\documentclass[sigconf, table]{acmart}
\AtBeginDocument{%
  \providecommand\BibTeX{{%
    \normalfont B\kern-0.5em{\scshape i\kern-0.25em b}\kern-0.8em\TeX}}}
 
\copyrightyear{2024} 
\acmYear{2024} 
\setcopyright{acmlicensed}\acmConference[CHI '24]{Proceedings of the CHI Conference on Human Factors in Computing Systems}{May 11--16, 2024}{Honolulu, HI, USA}
\acmBooktitle{Proceedings of the CHI Conference on Human Factors in Computing Systems (CHI '24), May 11--16, 2024, Honolulu, HI, USA}
\acmDOI{10.1145/3613904.3642509}
\acmISBN{979-8-4007-0330-0/24/05}

\usepackage{titlesec}
\usepackage{tabularx}
\usepackage{booktabs}
\usepackage{graphicx}
\usepackage{hyperref}
\usepackage{trimspaces}

 \usepackage{amsmath,amsfonts,amsthm}
\usepackage{booktabs}
\usepackage{tabularx}
\usepackage{geometry}

\usepackage{tabularray}
\usepackage{subcaption}
\usepackage{enumitem}

\usepackage{float} 
\usepackage{soul,color}
\usepackage[utf8]{inputenc}

\usepackage{longtable}
\usepackage[flushleft]{threeparttable}
\usepackage{chngcntr}
\usepackage{xcolor}
\definecolor{editCol}{rgb}{0.0, 0.0, 0.0}
\newcommand{\edit}[1]{{\textcolor{editCol}{#1}}}
\usepackage{pbox}
\usepackage{multirow}
\usepackage[colorinlistoftodos]{todonotes}
\usepackage{array}
\usepackage{makecell}
\usepackage{afterpage} 
\usepackage{hhline}
\usepackage[title]{appendix}
\usepackage{subcaption,graphicx}
\usepackage{ragged2e}

\newcommand{\oneS}{\ensuremath{{}^{\textstyle *}}}

\newcolumntype{L}[1]{>{\raggedright\let\newline\\\arraybackslash\hspace{0pt}}m{#1}}
\newcolumntype{C}[1]{>{\centering\let\newline\\\arraybackslash\hspace{0pt}}m{#1}}
\newcolumntype{R}[1]{>{\raggedleft\let\newline\\\arraybackslash\hspace{0pt}}m{#1}}

  {\list{}{\leftmargin=0.3in\rightmargin=0.3in}\item[]}%
  {\endlist}

\begin{document}

\title[Online Risk Experiences of LGBTQ+ Youth]{Examining the Unique Online Risk Experiences and Mental Health Outcomes of LGBTQ+ versus Heterosexual Youth}


\author{Tangila Islam Tanni}
\email{TangilaIslam.Tanni@ucf.edu}
\orcid{0000-0002-6737-7377}
\affiliation{%
  \institution{University of Central Florida}
  \city{Orlando}
  \state{Florida}
  \postcode{32826}
  \country{USA}
 }
 
\author{Mamtaj Akter}
\email{Mamtaj.Akter@vanderbilt.edu}
\orcid{0000-0002-5692-9252}
\affiliation{%
  \institution{Vanderbilt University}
  \city{Nashville}
  \state{Tennessee}
  \postcode{37212}
  \country{USA}
}

\author{Joshua Anderson}
\email{Jo178841@ucf.edu}
\orcid{0009-0007-0238-0165}
\affiliation{%
  \institution{University of Central Florida}
  \city{Orlando}
  \state{Florida}
  \postcode{32826}
  \country{USA}
 }
 
\author{Mary J. Amon}
\email{MJAmon@ucf.edu}
\orcid{0000-0003-0026-7568}
\affiliation{%
  \institution{University of Central Florida}
  \city{Orlando}
  \state{Florida}
  \postcode{32826}
  \country{USA}
 }
 
\author{Pamela J. Wisniewski}
\email{Pamela.Wisniewski@vanderbilt.edu}
\orcid{0000-0002-6223-1029}
\affiliation{%
  \institution{Vanderbilt University}
  \city{Nashville}
  \state{Tennessee}
  \postcode{37212}
  \country{USA}
}

\renewcommand{\shortauthors}{Tanni et al.}
\begin{abstract}
\edit{We collected and analyzed Instagram direct messages (DMs) from 173 youth aged 13–21 (including 86 LGBTQ+ youth). We examined youth's risk-flagged social media trace data with their self-reported mental health outcomes to examine how the differing online experiences of LGBTQ+ youth compare with their heterosexual counterparts.} We found that LGBTQ+ youth experienced significantly more high-risk online interactions compared to heterosexual youth. LGBTQ+ youth reported overall poorer mental health, with online harassment specifically amplifying Self-Harm and Injury. LGBTQ+ youth's mental well-being linked positively to sexual messages, unlike heterosexual youth. Qualitatively, we found that most of the risk-flagged messages of LGBTQ+ youth were sexually motivated; however, a silver lining was that they sought support for their sexual identity from peers on the platform. \edit{The study highlights the importance of tailored online safety and inclusive design for LGBTQ+ youth, with implications for CHI community advancements in fostering a supportive online environments.}

\textbf{Content Warning:} This research discusses sensitive topics, including explicit sexual content, abusive language, and homophobic slurs, which may cause discomfort. Reader discretion is advised.  

\end{abstract}



\maketitle

\section{Introduction}
Social media platforms like Instagram, TikTok, and Snapchat are popular among teens, especially those who feel isolated or marginalized, such as youth from the lesbian, gay, bisexual, transgender, queer or questioning communities (LGBTQ+) ~\cite{newport2021, aecf2021}. As of 2020, an estimated 9.5\% of the youth population in the United States is comprised of LGBTQ+ youth (ages 13-17)~\cite{williamsinstitute}, which represents a growing demographic of youth with unique needs related to their sexual and gender identities~\cite{steinke2017meeting}. Consequently, many LGBTQ+ youth turn to social media platforms as a means to seek emotional and social support, foster friendships, and find validation~\cite{mcinroy2020s}. Community engagement within the LGBTQ+ community can aid these youth in their identity exploration and foster confidence in themselves and their sexuality~\cite{mcinroy2019lgbtq+, socialbenefit}. However, other researchers have highlighted the negative effects of online engagement for this vulnerable community, as LGBTQ+ youth also face unique risks when it comes to navigating online spaces~\cite{abreu2018cyberbullying, schneider2012cyberbullying, zych2016cyberbullying, Hatchel_LGBTQYA_2021}. For instance, research has shown that they are disproportionately a target of harassment and other forms of cyberbullying, compared to their heterosexual counterparts~\cite{abreu2018cyberbullying}. These negative experiences have been shown to cause mental health issues for LGBTQ+ youth and, at the extreme, may even lead to suicidal tendencies or self-harm~\cite{lewis2011scope}. 

Recently, social media interactions have been trending towards private message communication instead of public postings due to several reasons, including privacy concerns~\cite{facebook2019, we2020, private2019}, to escape public scrutiny, and to reduce data visibility~\cite{simplilearn2021}. Private forms of communication have unique benefits (e.g., social support, advice from peers, etc.) but may also subject users to new harms (e.g., targeted harassment, hate speech, sexual grooming). Private conversations allow increased anonymity, sometimes facilitating inappropriate or uncharacteristic behaviors online~\cite{zimmerman2016online}. For instance, prior research suggests that most of the abuse on social media platforms comes in the form of Direct Messages (DMs) from strangers~\cite{twitter2019}. As DMs are considered private conversations, the majority of social media platforms do not address hate speech or bullying experienced through these channels~\cite{insta2021}. Therefore, these private chats represent a potentially unique threat yet to be examined, particularly in the context of LGBTQ+ youth in comparison with heterosexual youth. Yet, a common theme among existing research studies is that they have historically relied heavily on youth self-reports~\cite{almeida2009emotional} or publicly scraped social media data~\cite{saha2019language, rieger2021assessing}, which constrains what we know about LGBTQ+ youth's experiences in private spaces on social media. Therefore, to better understand the online risks encountered by LGBTQ+ youth in private social media spaces and how this may influence their overall mental health, the current study poses the following high-level research questions: 

 \begin{itemize}
 
    \item \textbf{RQ1:} \textit{Do LGBTQ+ youth experience more frequent and/or severe online risks in their private social media conversations compared to heterosexual youth?} 
    
    \item \textbf{RQ2:} \textit{Do LGBTQ+ youth who experience negative online experiences within their private conversations have more mental health challenges compared to heterosexual youth?} 
   
   \item \textbf{RQ3:} \textit{What is the unique nature of the online risks LGBTQ+ youth experience in private online chats?}
   
 \end{itemize}

To answer our research questions, we conducted a user study with 173 youth (ages 13-21), including 86 LGBTQ+ participants. We measured several pre-validated constructs regarding the youths' mental health and had them upload and flag their Instagram DMs that they felt uncomfortable or unsafe about. To answer RQ1, we first conducted between-group tests to evaluate LGBTQ+ youth versus heterosexual youth's self-reported online risk experiences and mental health challenges. For RQ2, we also applied regression models to analyze the relationship between online risk experiences and the mental health outcomes of youth participants. Finally, for RQ3, we conducted a qualitative analysis of the flagged messages for deeper insights into LGBTQ+ youth's online risk experiences.

\edit{Overall, we found that the private online experiences of LGBTQ+ youth differ significantly from the experiences of heterosexual youth in several key ways. For RQ1, we found that while LGBTQ+ youth do not report more online risk experiences overall, the experiences that they report are of higher risk severity and are more likely to involve messages that contain sexual or self-injurious content. For RQ2, we confirmed findings from prior literature that found LGBTQ+ youth report significantly worse mental health outcomes (e.g., Self-Harm and Injury, Depression, and decreased Mental Well-being) compared to heterosexual youth. In addition, we uncovered two significant interaction effects; LGBTQ+ youth who receiving more online harassment report significantly more self-harm and self-injurious behavior, while this effect was not seen among heterosexual youth. In contrast, LGBTQ+ youth who received more sexually risky messages reported higher levels of mental well-being, while this effect was in the opposite direction for heterosexual youth. Our RQ3 qualitative analyses revealed additional insights into our quantitative findings. The conversations flagged as risky by LGBTQ+ youth often contained sexual undertones, even when the intent was to harass. However, some of these risky messages gave LGTBQ+ youth the opportunity to explore their sexual and gender identities. Importantly, Instagram DMs gave them a channel to share about their struggles with others who provided needed support. Our results provide useful insights for supporting the unique online experiences of LGBTQ+ youth, which we unpack in our discussion.}

\edit{Our research makes a distinct contribution to the CHI community, particularly in the context of inclusive design and research advocating for LGBTQ+ populations. Building upon the LGBTQ+ advocacy research within the CHI community (c.f.,  ~\cite{acena2021my,gatehouse2018troubling,carrasco2018queer}), we advance this body of knowledge by delving into youths' private social media experiences through the use of their Instagram Direct Messages. Focusing on this timely subject, our research furthers understandings of the challenges faced by a particularly vulnerable demographic, LGBTQ+ youth, in the realm of online threats and bullying experiences but also critically evaluates the efficacy of existing filtering tools and protective measures for vulnerable users in the face of cyberbullying, urging a call-to-action for risk prevention and mitigation program. The broader implications of the research extend to the advancement of Human-Centric Machine Learning (HCML) algorithms designed to detect and address online sexual risks, thus contributing significantly to the ongoing efforts to enhance LGBTQ+ youth’s digital safety and well-being. Specifically, we make the following unique research contributions:
\begin{itemize}
    \item Through a mixed method study, we disentangled how LGBTQ+ and heterogeneous youth's  online risk experiences differed, finding that LGBTQ+ youth report higher risk conversations containing sexual and self-harm content.
    \item Our research highlighted a range of online risks experienced by LGBTQ+ youth and demonstrates how the multi-faceted risk factors go beyond online public spaces and occur in private contexts.
    \item We examined social media direct message data to underscore the association between receiving harassing messages and the self-harming behaviors of LGBTQ+ youth.
    \item We gained deeper understanding of how LGBTQ+ youth did not only experience sexual harassment and unsolicited sexual content from strangers and peers, they also sought support from their peers. 
\end{itemize}
}


\textbf{Content Warning:} This study discusses delicate subjects such as sexual content, offensive language, and derogatory remarks aimed at LGBTQ+ individuals, which might evoke unease. Caution is advised for readers.  

\section{Related Work}
In this section, we synthesize related work regarding LGBTQ+ youth and their social media usage. Next, we introduce literature highlighting how LGBTQ+ youth are more vulnerable to online risks and how the negative online experiences impact mental health.

\subsection{LGBTQ+ Youth, Social Media, and Support}

LGTBQ+ turn to online platforms to seek information related to gender, sexuality, and romantic relationships, as well as to seek support for physical, mental, and sexual health needs~\cite{craig2014you,dehaan2013interplay,harper2009role, razi_let_2020,lucero2017safe,pyle2021lgbtq}. Yet, LGBTQ+ youth reported having smaller online social networks compared to heterosexual youth~\cite{charmaraman2021young}. In contrast, they reported joining online communities or groups more frequently to feel less isolated or lessen social isolation compared to heterosexual youth~\cite{charmaraman2021young}. Furthermore, the likelihood of LGBTQ+ teenagers having friends they only know online is higher as they often view online friends as being more supportive than their in-person peers~\cite{ybarra2015online}.

Past research highlighted the role of online social networks as safe havens for LGBTQ+ youth who live in hostile environments. They found that the anonymity and lack of geographic restrictions in digital spaces offer an ideal platform for coming out, engaging with a communal gay culture, experimenting with non-heterosexual intimacy, and socializing with other LGBTQ+ youth~\cite{hillier2007building}. Furthermore, LGBTQ+ youth also interact and build connections with a specific community and engage themselves with content that validates and recognizes their unique experiences with being an LGBTQ+ individual. However, seeking social support in online communities often requires disclosing personally sensitive information, such as one's gender identity and personal struggles~\cite{haimson2015disclosure,hillier2007building}. In the early phases of developing their LGBTQ+ identities, LGBTQ+ individuals may use social media platforms as informal learning environments~\cite{fox2016queer}. For example, previous research findings identified three educational purposes associated with online information seeking on social media: traditional learning about LGBTQ+-related issues, social learning involving mirroring role models or other LGBTQ+ people's behavior, and experiential learning with online dating sites and dating apps. These learning behaviours were common during the coming-out process~\cite{fox2016queer}. 

Decisions of coming out are further complicated by the potential risks and negative consequences of disclosing one's LGBTQ+ identity in unsupportive environments~\cite{devito2018too,haimson2015disclosure}. Violations of privacy~\cite{geeng2021lgbtq}, being disowned by family members~\cite{cho2018default}, and rejection from society are potential consequences of coming out. Regardless of these fears, youth often choose to disclose their sexual identity online to obtain needed support and social connections~\cite{geeng2021lgbtq,fox2015queer}. Coming out about one's sexual identity is positively associated with relationship satisfaction, self-efficacy, self-esteem, lower anxiety, and lower levels of depression ~\cite{mustanski2011mental,meyer2019impact,andalibi2019happens}. Thus, LGBTQ+ youth have multiple co-acting motives for seeking online support and also experience a range of benefits and risks within online spaces. However, most prior research in this area has focused on how LGBTQ+ youth self-report on their experiences in reference to online public spaces (e.g., via surveys, focus group studies, interviews, or analyzing public social media data). Further research is required to understand the online risks in private one-to-one communication, particularly through the analysis of real-world direct messages.

\subsection{LGBTQ+ Youth Online Risks and Well-being}

Although the Internet provides LGBTQ+ youth numerous opportunities to form new relationships and explore their sexual identities~\cite{geeng2021lgbtq,fox2015queer}, it can pose a range of online risks, such as privacy violations~\cite{pink2021,egypt2017,bbc2021}, cyberbullying~\cite{ybarra2015online,schneider2012cyberbullying,li2020trends}, sexual risks~\cite{o2011impact}, and other types of abuse~\cite{powell2020digital}. Social media platforms provide social visibility, connectivity, feedback, and ease of accessibility, which can increase cyberbullying compared to traditional bullying~\cite{fox2015dark}. In particular, LGBTQ+ youth experience higher rates of cyberbullying (\cite{schneider2012cyberbullying,ybarra2015online}) and are often targeted due to their nonconforming sexual identities~\cite{aoyama2011cyberbullying,abreu2018cyberbullying}. Moreover, a greater number of polyvictimization incidents, where youth who experience one type of victimization are more at risk of experiencing other types as well, are prevalent among LGBTQ+ youth~\cite{finkelhor2007polyvictimization, finkelhor2009pathways,turner2010poly,sterzing2017social}. LGBTQ+ youth report being bullied or harassed online nearly three times as frequently as non-LGBTQ+ youth (42\% vs. 15\%)~\cite{gay_out_2013}. Other forms of online risks, such as hate-based abuse, violence, and discrimination are also prevalent among sexual minority youth~\cite{guasp2013homophobic,hiller2005writing}. 

As a result, LGBTQ+ youth experience "minority stress," a chronic form of stress caused by negative social experiences associated with identifying with a minority group~\cite{aoyama2011cyberbullying,meyer2003prejudice,saha2019language}, which can be amplified in social media contexts~\cite{saha2019language}. According to the minority stress theory, sexual and gender minority health disparities may be traced to stressors brought on by hostile, homophobic, and transphobic cultures. These stressors frequently result in harassment, abuse, and victimization and may ultimately affect access to care~\cite{saha2019language,denton2014stigma,frost2015minority,hendricks2012conceptual}. In particular, stigma, prejudice, and discrimination create a hostile social environment that can lead to mental health difficulties~\cite{meyer1995minority}. LGBTQ+ youth reported higher rates of suicidal thoughts and depressive symptoms than their straight peers, which are influenced by negative experiences, including discrimination and victimization~\cite{marshal2011suicidality}. These mental health issues, which result from bullying and other online risk factors, can significantly impact well-being and relationships with others~\cite{keighley2022hate}. 

Whereas much of the current research is focused on young LGBTQ+ adults, more research is needed to understand how younger LGBTQ+ adolescents (ages 13-17) experience these online risk factors. This is especially true given that the permissible age for having a social media account (e.g., for Facebook or Instagram) is 13~\cite{holloway2016internet, montgomery2017ensuring}, a time during which LGBTQ+ youth may be particularly susceptible to online dangers and associated mental health consequences. Moreover, the risk factors associated with online engagement highlight the critical need to go beyond just examining public online communities, but also understanding the multi-faceted risk factors that occur in private contexts as well. As such, our research complements and extends beyond previous work by examining the online risk experiences and associated mental health outcomes of adolescent LGBTQ+ youth between the ages of 13 to 21. Our study adds depth by adopting a mixed-method approach and pairing the self-reported mental health data of LGBTQ+ youth with their real-world social media private conversations that they flagged as unsafe or risky. By triangulating these two data sources, our analysis is one of the first to deeply examine the association between these two facets of LGBTQ+ youths' personal and social experiences, contributing to a more holistic representation of LGBTQ+ youth's online experiences.

\section{Methods}
In this section, we provide a detailed description of our study, including the survey design, Instagram data collection procedures, and the data analysis approach we used to answer each of our research questions.

\subsection{Study Overview}
We conducted a user study of youth (ages 13-21), who first completed a web-based survey, then were asked to upload their Instagram data and subsequently flag their DM conversations for interactions that made them feel uncomfortable or unsafe. We chose Instagram as our social media platform of choice because, according to Pew Research, 72\% of teens use Instagram, making it one of the most popular social media platforms among youth~\cite{anderson2018teens}. We recruited participants, who met the following eligibility criteria: 1) English speakers based in the United States between the ages of 13-21, 2) Had an active Instagram account at while they were a teen (ages 13-17), 3) Communicated with at least 15 people through Instagram direct messaging (DMs), and 4) Had at least two conversations with other users that made them feel unsafe or uncomfortable. If participants met this eligibility criteria, parental consent was obtained for those under 18; otherwise, participants consented to participate themselves. Participants were compensated with a \$50 Amazon gift card for their time and data. This study was approved by our university's Institutional Review Board (IRB).

\subsection{Ethical Considerations}
We took the utmost care to protect the participants' anonymity and privacy due to the complicated and sensitive nature of the data collected. First, we obtained IRB approval for our work, declared our position as mandated child abuse reporters within the statement of informed consent in the case of an impending danger posed to a minor, and followed the guidance of Badillo-Urquiola et al.~\cite{badillo_conducting_2021} on conducting risky research with minors. For instance, we provided explicit instructions NOT to share digital images that depicted the nudity of a minor and outlined our statutory responsibility to report child pornography to the proper authorities. We also provided clear directions to delete such materials prior to data sharing. Additionally, we secured a Certificate of Confidentiality from the National Institute of Health, which further protected participants by precluding the subpoenaing of the data during legal discovery. To further protect the privacy of our participants and everyone else who participated in direct message conversations, we removed all personally identifiable information from any textual data reported in our results and paraphrased all quotations. To prevent sharing the data to outside parties, we also did not analyze our data using any cloud-based services. Researchers were prohibited from downloading the data onto personal devices and were required to complete IRB Human Subjects CITI training. We also gave research assistants, who assisted in verifying and qualitatively analyzing the data, mental health support, and adequate breaks.

\begin{table}[!hbt]
 \centering
  \caption{Summary of participants demographics}
  \label{demographic 1}
  \begin{tabular}{lccc}
    \toprule
   & \multicolumn{1}{c}{\textit{n} (\%)}\\
    Total (\textit{N}) & 173 (100\%)\\
    \midrule
    Age (M (SD)) & 17 (2.14)\\
    \midrule
    Gender Identity & \\
    \hspace{1 cm} Female  & \multicolumn{1}{l}{117 (67.63\%)}\\
    \hspace{1 cm} Male  & \multicolumn{1}{l}{39 (22.54\%)}\\
    \hspace{1 cm} Non-binary  & \multicolumn{1}{l}{14 (8.09\%)}\\
    \hspace{1 cm} Prefer to self-identify  & \multicolumn{1}{l}{3 (1.73\%)}\\
    \midrule
    Race & \\
    \hspace{1 cm} White/Caucasian  & \multicolumn{1}{l}{92 (53.18\%)}\\
    \hspace{1 cm} Black/African-American  & \multicolumn{1}{l}{45 (26.01\%)}\\
    \hspace{1 cm} Asian or Pacific Islander  & \multicolumn{1}{l}{36 (20.81\%)}\\
    \hspace{1 cm} Hispanic/Latino  & \multicolumn{1}{l}{26 (15.02\%)}\\
    \hspace{1 cm} American Indian/Alaska Native & \multicolumn{1}{l}{7 (4.04\%)}\\
    \hspace{1 cm} Prefer to self-identify& \multicolumn{1}{l}{5 (2.89\%)}\\
    \hspace{1 cm} Two or more races or ethnicity & \multicolumn{1}{l}{32 (18.49\%)}\\
    \midrule
    Sexual Identity & \\
    
    \hspace{1 cm} Heterosexual/straight & 87 (50.29\%)\\
    \hspace{1 cm} Bisexual & 47 (27.16\%)\\
    \hspace{1 cm} Homosexual/gay & 17 (9.82\%)\\
    \hspace{1 cm} Self-identify & 22 (12.71\%)\\
     \hspace{2 cm} Pansexual  & \multicolumn{1}{l}{10 (45.45\%)}\\
     \hspace{2 cm} Omnisexual & \multicolumn{1}{l}{2 (9.00\%)}\\
     \hspace{2 cm} Asexual & \multicolumn{1}{l}{2 (9.00\%)}\\
     \hspace{2 cm} Biromantic asexual & \multicolumn{1}{l}{1 (4.55\%)}\\
     \hspace{2 cm} Neptunic & \multicolumn{1}{l}{1 (4.55\%)}\\
     \hspace{2 cm} Panoromantic and demisexual & \multicolumn{1}{l}{1 (4.55\%)}\\
     \hspace{2 cm} Fluid & \multicolumn{1}{l}{1 (4.55\%)}\\
     \hspace{2 cm} Curious & \multicolumn{1}{l}{1 (4.55\%)}\\
     \hspace{2 cm} I don't know & \multicolumn{1}{l}{3 (13.64\%)}\\
   
  \bottomrule

\end{tabular}
\end{table}

\subsection{Participant Recruitment and Demographics}
A total of (\textit{N} = 173) participants completed the study. In the survey, participants were asked to report several demographic characteristics, including their gender identity and sexual identity (e.g., Heterosexual or straight, Homosexual or gay, Bisexual, or Prefer to self-identify). Details of participant demographics based on age, gender identity, race, and sexual identity can be found in Table \ref{demographic 1}. Since our participants could select multiple races, the total percentages of all categories can be greater than 100\%. We grouped homosexual/gay, bisexual, and individuals who preferred to self-identify as LGBTQ+ youth, as the purpose of the term "LGBTQ+" is to be inclusive of all same-gender attracted and trans people~\cite{lgbtq2020}. Among all participants who completed the study, 50.29\% (\textit{n} = 87) of youth identified as heterosexual, whereas 49.71\% (\textit{n} = 86) were LGBTQ+. Participants also responded to the social media usage questions, which helped us better understand their Instagram use. For almost half (49.71\%) of our participants, Instagram was their most used social media platform with 64.16\% (\textit{n} = 111) participants having more than one Instagram account. 47.97\% (\textit{n} = 83) of the participants used Instagram several times a day, 24.27\% (\textit{n} = 42) used every day or almost every day, 21.96\% (\textit{n} = 38) used several times an hour, 4.04\% (\textit{n} = 7) used once or twice a week, 1.15\% (\textit{n} = 2) used less than once a month, and 0.57\% (\textit{n} = 1) used less than once a week. 97.68\% (\textit{n} = 169) participants never met their Instagram followers in person. 

\subsection{Data Collection and Risk-Flagging} 
After completing the web-based survey, participants uploaded their Instagram data to a secure web-based system~\cite{razi2022instagram}. \edit{This web-based system was developed using PHP with an Amazon Web Services (AWS) back-end infrastructure to encrypt and store the data. Participants requested their data from Instagram and uploaded it in the form of JSON files that were translated and stored on our AWS server. Then, participant's} DMs were presented \edit{through our web-based interface} in reverse chronological order for them to review and flag the uncomfortable or unsafe (i.e., “risky”) messages of each conversation. Participants marked individual messages by risk type (e.g., sexual content, harassment, spam, hate speech, violence, illicit actions, and self-injury) and risk level (low, medium, high). While aligned with Instagram’s reporting features~\cite{facebook2022}, participants could denote their own risk type. We provided a benchmark for risk level: low for discomfort without harm, medium for potential emotional or physical harm, and high for actual harm. The dataset included 32,055 DM conversations with over 6 million (6,863,161)  messages. Participants marked 2,515 conversations as 'risky,' flagging a total of 3,023 messages by risk type and level. High-level risks comprised 402 messages, while medium and low risks were 821 and 1,813, respectively. Table \ref{severity_youth} displays message counts based on risk types and severity. Past studies (as synthesized in section 2.2) suggested that LGBTQ+ youths face bullying and harassment more on both offline and online, including on social media platforms. Based on prior findings from the literature, we hypothesized that: \\

\textbf{H1:}  \textit{LGBTQ+ youth will report encountering more online risks than heterosexual youth.} \vspace{1mm}

\textbf{H2:}  \textit{LGBTQ+ youth will report encountering more high-risk online experiences than heterosexual youth.}\vspace{1mm}\\\

Given that the literature was not conclusive regarding the specific types of online risks LGBTQ+ youth encountered compared to their heterosexual counterparts, we examined these relationships in our results without formalizing hypotheses based on risk types.

\subsection{Mental Health and Well-being Measures}
In the survey, participants were also asked to report on several pre-validated constructs to assess their mental health and well-being. These survey constructs included: The Short Warwick-Edinburgh Mental Well-Being Scale (SWEMWBS)~\cite{wellness}, the Patient Health Questionnaire (PHQ)-9, a commonly used measure for Depression~\cite{phq-9_2001}, and the Inventory of Statements about Self-injury (ISAS)~\cite{harm}. These constructs were measured on a 5-point Likert scale. \edit{To ensure alignment with other survey measures, we modified the (PHQ)-9 scale from a 4 to a 5-point scale. This adjustment was validated through factor analyses, incorporating Horn's parallel analysis in exploratory factor analysis (\cite{horighausen2015}) and taking into account prior research on the 4-point scale~\cite{vu2022factor}. The subsequent confirmatory factor analysis produced favorable fit indices, surpassing the 0.9 threshold. Notably, both RMSEA and SRMR scores exceeded the recommended cutoff of 0.05, confirming the validity of our refined 5-point (PHQ)-9~\cite{cfa2020, goldberg1992development, kline2023principles}}. We tested for internal consistency using Cronbach's alpha and all values were above the acceptable threshold (0.7)~\cite{cho2015cronbach}. Table ~\ref{Mental Health Comparsion} shows the scale reliability and statistical description of the mental health constructs for both LGBTQ+ and heterosexual youth who participated in our study. As discussed in the prior literature, many studies have found evidence that LGBTQ+ youth showed greater depressive symptoms and, in some cases, suicidal ideation as a result of the negative consequences of stigma, prejudices, discrimination, and abusive attitudes held by the predominately heterosexual society~\cite{clarke2010lesbian,yi2017health,stojisavljevic2017devil}. Generally, LGBTQ+ youth are found to have poor mental health conditions due to the negative experiences faced in both offline and online settings. 
Based on this literature, we formulated the following hypotheses related to the well-being and mental health of LGBTQ+ versus hetersexual youth: \\

\textbf{H3:}  \textit{LGBTQ+ youth will report worse mental health outcomes, including a) lower Well-Being, b) higher Depression, c) and higher Self-Harm and Injury than heterosexual youth.} \vspace{1mm}

\textbf{H4:}  \textit{Youth who experience more high-risk online experiences will report worse mental health outcomes than those who report experiencing more low-level online risk.} \vspace{1mm}

\textbf{H5:}  \textit{The negative mental health outcomes associated with online risk experiences will be moderated by sexual identity, such that the negative effect will be stronger for LGBTQ+ youth than for heterosexual youth.} \\

In the next section, we describe our statistical methods for testing our research hypotheses, as well as our qualitative approach for gaining additional insights.

\subsection{Data Analysis Approach}
In this section, we first describe how we conducted our statistical analyses to answer our hypotheses related to RQ1 and RQ2. Then, we describe our qualitative analysis approach for investigating RQ3.

\subsubsection{Examining the Differing Online Risk Experiences of LGBTQ+ Versus Heterosexual Youth (RQ1).}  
To investigate differences in online risk experiences encountered by LGBTQ+ and heterosexual youth, we first analyzed the risk-flagged messages provided by youth. Among the total (\textit{n=3,023}) messages, 48.56\% (\textit{n=1468}) were flagged by LGBTQ+ youth, while 51.44\% (\textit{n=1555}) were flagged by heterosexual youth. Table \ref{severity_youth} summarizes the number and percent of risk-flagged messages by risk type and level for both LGBTQ+ and heterosexual youth. We conducted a between-group chi-square test ($\chi^2$) to compare the difference in total number of online risks experienced by these two youth groups (H1). The $\chi^2$ test of independence is employed for between-group comparisons involving two or more groups~\cite{mchugh2013chi}, and previous studies contrasting sexual minority groups with heterosexual individuals have successfully employed this approach~\cite{otherchi}. 

Furthermore, we explored significant between-group differences between LGBTQ+ and heterosexual youth based on risk level to test H2. Specifically, we utilized standardized residuals to provide insights into cells significantly impacting the chi-square value. Cells with standardized residuals exceeding +2 are considered major contributors, while those surpassing -2 weakly contribute to the overall chi-square calculation~\cite{chi}. We performed another $\chi^2$ test to identify significant differences in the total number of messages falling into each of the six risk types, as self-annotated by the youth. Similar to the approach with standardized residuals described above, we employed this analysis to gain insights into risk type disparities within the dataset.

\begin{table*}[!h]
  \footnotesize
   \centering
\caption{Counts and percentage of risks based on risk type and severity for LGBTQ+ and heterosexual participants (\textit{N} = 173)}
\label{severity_youth}
\begin{center}
\begin{tabular}{| c | c | c | c | c | c | c |}
\hline
\multirow{2}{*}{\textbf{Risk Type}} &
  \multicolumn{3}{c|} {\textbf{LGBTQ+} (\textit{N} = 86)}  &
  \multicolumn{3}{c|} {\textbf{Heterosexual} (\textit{N} = 87)} \\
\cline{2-7}
 & \textbf{Low} & \textbf{Medium} & \textbf{High} & \textbf{Low} & \textbf{Medium} & \textbf{High} \\
\hline

Sexual messages & \textit{n} = 286 & \textit{n} = 171 & \textit{n} = 141 & \textit{n} = 339 & \textit{n} = 151 & \textit{n} = 69 \\
\textit{N} = 1157 & 35.18\% & 43.85\% & 53.21\% & 34.35\% & 35.03\% & 50.36\% \\ \hline
Harassment & \textit{n} = 276 & \textit{n} = 101 & \textit{n} = 64 & \textit{n} = 327 & \textit{n} = 159 & \textit{n} = 36 \\
\textit{N} = 963 & 33.94\% & 25.89\% & 24.15\% & 33.13\% & 36.89\% & 26.27\% \\ \hline
Spam and others & \textit{n} = 158 & \textit{n} = 39 & \textit{n} = 9 & \textit{n} = 185 & \textit{n} = 52 & \textit{n} = 13 \\
\textit{N} = 456 & 19.43\% & 10.00\% & 3.40\% & 18.74\% & 12.06\% & 9.49\% \\ \hline
Hate Speech and violence & \textit{n} = 48 & \textit{n} = 47 & \textit{n} = 28 & \textit{n} = 68 & \textit{n} = 41 & \textit{n} = 12 \\
\textit{N} = 244 & 5.90\% & 12.05\% & 10.57\% & 6.89\% & 9.51\% & 8.76\% \\ \hline
Illicit actions & \textit{n} =30 & \textit{n} = 21 & \textit{n} = 13 & \textit{n} = 55 & \textit{n} = 21 & \textit{n} = 7 \\
\textit{N} = 147 & 3.69\% & 5.38\% & 4.90\% & 5.57\% & 4.87\% & 5.10\% \\ \hline
Self-injury & \textit{n} = 15 & \textit{n} = 11 & \textit{n} = 10 & \textit{n} = 13 & \textit{n} = 7 & \textit{n} = 0 \\
\textit{N} = 56 & 1.84\% & 2.82\% & 3.77\% & 1.31\% & 1.62\% & 0.00\% \\ \hline
Total & \textit{N} = 813 (100\%) & \textit{N} = 390 (100\%) & \textit{N} = 265 (100\%) & \textit{N} = 987 (100\%) & \textit{N} = 431 (100\%) & \textit{N} = 137 (100\%) \\
\hline

\end{tabular}
\end{center}
\vspace{-4mm}
\end{table*}

\subsubsection{Exploring Associations Between Mental Health and Negative Online Experiences (RQ2).} 
To operationalize the frequency and levels of risk for each participant, we utilized a weighted model, with higher severity risks having a higher weight. \edit{Our methodology involved multiplying the number of messages within each risk type by the corresponding risk weight. This resulted in a comprehensive risk score for each risk type and user, providing a nuanced representation of the interplay between message volume and risk severity.} Next, we averaged items associated with the three mental health constructs and calculated the mental health scores for each participant. After calculating their Self-Harm and Injury, Depression, and Mental Well-Being scores for both LGBTQ+ and heterosexual youth, we examined the distribution and variance of the two groups for each construct. As the two populations were normally distributed, and the variances for the constructs were not equal for our independent populations, we performed Welch's two-sample \emph{t}-test to compare the between-group differences in their mental health scores \cite{welches}. 

Next, we investigated the relationship between the risky messages received via DMs and the mental health of these youth via multiple linear regressions, and we also tested whether this relationship is modulated by the self-reported sexual identity of these youth. We started by fitting our data into a linear regression model with each youth's risk type scores and sexual identity as predictor variables and mental health constructs as our outcome variables. Six risk-type scores (e.g., sexual message, harassment, spam and others, hate speech and violence, illicit actions, and self-injury) were entered as predictor variables. Three mental health scores (Depression, Self-Harm and Injury, Mental Well-Being) were the outcome variables. Furthermore, the sexual identity of these youth was used as an interaction term in each regression model. Here, messages categorized as self-injury are not synonymous with the mental health construct known as Self-Harm and Injury.
\begin{table*}[!h]
    \centering
\footnotesize
\caption{Codebook for Risk-Flagged Conversations of LGBTQ+ Youth (RQ3)}
\label{tab:codebook}
\begin{tabular}{ | p{4cm} | p{5.5cm} | p{7cm} |}
         \hline
   \textbf{Themes}  & \textbf{Subthemes (Codes)} & \textbf{Illustrative Quotations} \\ \hline
     \multirow{4}{4cm}[-2pt]{\vspace{0pt}\textbf{Most risky conversations had sexual undertones} \\
     } &   \vspace{0pt}\textbf{Exchanged sexual message content} \newline (Sexual content from strangers; from peers; Sent to others; Added to porn groupchats by strangers) &  
\vspace{0pt}
\edit{
\textbf{OP*:} \textit{hurry up bitch} \newline
\textbf{OP:} \textit{u wasting my time} \newline
 \textbf{P*:} \textit{[User ID] sent an attachment} \newline
 \textbf{OP:} \textit{oh girl, u are a good submissive. I fucking love u}} \newline
     \\ 
      \cline {2-3} 
      & \vspace{0pt}\textbf{Received sexual solicitations} \newline (Sexual solicitations from strangers; From peers; With monetary offers; Received persistent messages; Led to harassment)& 
\vspace{0pt}
\edit{
\textbf{OP:} \textit{Can I get my duck sucked} \newline
\textbf{P:} \textit{I don’t know you do I? And not by me }} \newline
      \\ 
      \cline {2-3}
      &  \vspace{0pt}\textbf{Got harassed for sexual identity} \newline (Harassed for LGBTQ+ identity by peers; by strangers) &  
\vspace{0pt}\edit{
\textbf{OP:} \textit{Faggots like you are in this world will take us in hell} \newline
\textbf{P:} \textit{leave me alone please}} \newline
      \\ 
      \cline {2-3}
       &  \vspace{0pt}\textbf{Received spam targeted to sexual identity} \newline (Advertisements; Scams targeted to LGBTQ+) & 
\vspace{0pt}\edit{
       \textbf{OP:} \textit{Hey, i’m from [Organization name]! I hope you’re having a great January. We are looking for brand ambassadors and reps. Please send us a message at our main account [UserID] ASAP! Have a wonderful day}} \newline
       \\ 
      \cline {2-3}
       \hline
      \multirow{2}{4cm}[2pt]{
      \newline
      \textbf{LGBTQ+ youth encountered general harassment and scams}
      } &  \vspace{0pt}\textbf{Got harassed} \newline (Harassed by strangers; By peers) &   
\vspace{0pt}\edit{
\textbf{OP:} \textit{Instagram User sent an attachment} \newline
\textbf{OP:} \textit{can we be friends} \newline
\textbf{OP:} \textit{Instagram User sent an attachment} \newline
\textbf{OP:} \textit{bitch talk with me}} \newline \\
     \cline {2-3}
          &  \vspace{0pt}\textbf{Received spam/scams} \newline (Received general advertisements; scams; ads for illegal substances)&  
\vspace{0pt}\edit{
\textbf{OP:} \textit{you won’t believe this, I just got sent a \$1,000 GiftCardVisa yesterday and all I had to do was participate a little and they sent me it. I used it to pick up a new gaming computer Ive been eyeing LOL. I thought u would want one since your my follower. Hurry thought there is just a few left!! Go}}  \newline
   \\ 
       
       \hline
     \multirow{3}{4cm}[-7pt]{
     \textbf{LGBTQ+ youth received support from peers regarding negative  experiences} 
     }  &  \vspace{0pt}\textbf{Shared negative experiences related to sexual identity} \newline (Received support for negative experiences with trans-phobic people; Fear of coming out; Urge to self-harm) & 
\vspace{0pt}\edit{
\textbf{P:} \textit{I am tired of my dad. He gives me pills to cure my homosexuality. Its as if a disease my parents want me to get well. At this point I don’t think I should be living anymore.}
\textbf{OP:} \textit{I am sorry your going through these. It suckz when family is liek them. Can you move out}} \newline
     \\ 
     \cline {2-3}
     &  \vspace{0pt}\textbf{Shared mental health concerns} \newline (Received support for sad feelings; Depressive thoughts) & 
\vspace{0pt}\edit{
\textbf{P:} \textit{ I ve bn having break up after break up. this time her problem was my bisexuality. I feel like cuting, I tried so many times} \newline
\textbf{OP:} \textit{Honestly, she doesn’t deserve you and your such a great guy and any girl would be lucky to have you}} \newline
     \\ 
     \cline {2-3}
     & \vspace{0pt} \textbf{Received other support and advice} \newline (Received advice for general negative experiences; Drug abuse) &  
\vspace{0pt}\edit{
\textbf{OP:} \textit{Oh so u out for smoking blunt again?} \newline
\textbf{P:} \textit{a little} \newline
\textbf{OP:} \textit{u know u shouldn’t be smoking}} \newline
     \\
     \hline
 \multicolumn{3}{l}{ \edit{\textbf{*P:} Participant; \textbf{*OP:} Other Person}} \\
    \end{tabular}
    \label{tab:my_label}
\end{table*}

\subsubsection{Qualitative Analysis of Instagram Private Message Conversations (RQ3).}
To further explore the types of risks that LGBTQ+ youth faced in private messaging contexts, we performed a thematic analysis ~\cite{braun2012thematic} of the private message conversations risk-flagged by LGBTQ+ youth. Each conversation represented a messaging thread where the participant and other person/people messaged back and forth intermittently. Therefore, each conversation could have either a single message or a series of messages that were exchanged over a long period of time. There were a total 1,468 unsafe messages (of 808 conversations) that were flagged by our LGBTQ+ youth participants. However, we analyzed all messages (N = 216,332) of these 808 conversations. All messages were in textual format. When participants or their conversation partners shared media, our data showed the message as "Instagram User sent an attachment." or "[Name] sent an attachment" instead of showing the actual file. In such cases, we used contextual cues from the larger conversation to interpret the risk. 

\edit{To complete the qualitative analysis, the second author initially familiarized themselves with the data by reading through the conversations and creating initial codes, which were then discussed among all co-authors. Subsequently, the second and last author worked together to iteratively establish consensus and incorporate codes as they emerged. They jointly collaborated to conceptually group the codes into cohesive subthemes and overarching themes. Our thematic analysis, as presented in \autoref{tab:codebook}, aimed to identify key characteristics of the risk-flagged conversations LGBTQ+ youth had with others.} For each theme, the total count and percentages of codes can be greater than the total number of conversations as we double-coded the conversations for different risk types. For example, there were instances when sexual message content, sexual solicitations, and bullying were all present in the same conversations.

\section{Results}
\edit{In the following sections, we highlight distinctions in the total number of online risks and their severity between LGBTQ+ and heterosexual youth by presenting the outcomes of chi-square tests. Subsequently, we delve into the results of linear regression models, investigating the correlation between online risk experiences and mental health outcomes for both groups. Lastly, we provide qualitative insights to enhance the overall understanding of online risks in private conversations among LGBTQ+ youth.}

\subsection{LGBTQ+ Youth at Higher Risk for Sexual and Self-Injury Messages (RQ1)}

\subsubsection{Flagged Messages by Risk Level (H1 \& H2).}
We first tested the degree to which LGBTQ+ youth flagged riskier messages compared to heterosexual youth. Overall, LGBTQ+ youth flagged (\textit{N = 1468}) messages, whereas heterosexual youth flagged (\textit{N = 1555}) as risky. Our chi-square test indicated significant variation in the total number of online risks experienced by LGBTQ+ youth versus heterosexual youth, $\chi^2$ (2, \textit{N} = 3023) = 57.167, \textit{p} < .001. However, this effect was in the opposite direction than hypothesized. Hence, the result did not support our first hypothesis (H1) that LGBTQ+ youth would report more online risks than heterosexual youth. In contrast, heterosexual youth reported significantly more online risks than LGBTQ+ youth.

\begin{figure*}
    \centering
\begin{subfigure}[t]{.45\textwidth}\centering
  \includegraphics[width=\columnwidth]{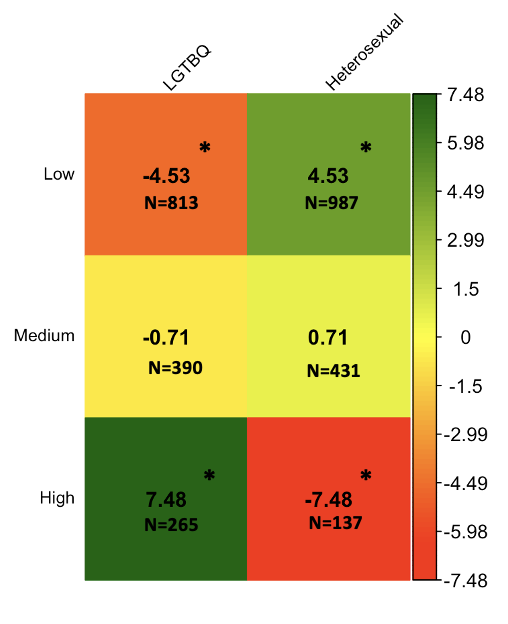}
  \caption{}
\end{subfigure}%
\hspace{1em}
\begin{subfigure}[t]{.45\textwidth}\centering
  \includegraphics[width=\columnwidth]{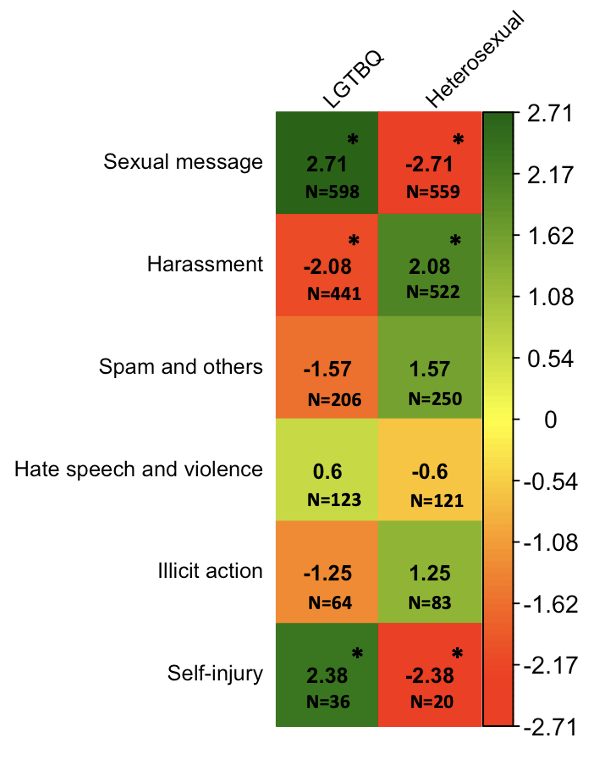}
  \caption{}
\end{subfigure}
\caption{Results (Standardized residuals) from chi-square test showing the between-group analysis of online risk messages encountered by LGBTQ+ and heterosexual youth based on (a) risk level and (b) risk types (\textit{N} = 3023). ($^*$) indicates significant association. Note that green indicates a positive association, while red indicates a negative one.}~\label{chi_square}
\end{figure*}

For H2, Figure \ref{chi_square} (a) reveals that LGBTQ+ youth experienced a significantly higher frequency of high severity online risks. Moreover, a significant difference exists between heterosexual and LGBTQ+ youth regarding the number of low-level risk conversations, with heterosexual youth reporting more low-level risk conversations. However, there was no significant difference in medium-level risks based on sexual identity. Thus, these results support our hypothesis (H2).

\subsubsection{Flagged Messages by Risk Type.}
We conducted an additional chi-square test to identify differences between LGBTQ+ and heterosexual youth groups based on risk types. The chi-square test revealed significant differences between the two groups concerning risk types, $\chi^2$ (5, \textit{N} = 3023) = 16.927, \textit{p} = .004. Standardized residuals indicated a significant difference between LGBTQ+ and heterosexual individuals, particularly when examining messages containing sexual content, harassment, and self-injury. As shown in Figure \ref{chi_square} (b), LGBTQ+ youth experienced a significantly higher frequency of sexual messages and messages containing self-injury language compared to their heterosexual counterparts. In contrast, heterosexual youth reported a significantly higher number of harassment messages than LGBTQ+ youth. In the next section, we examine the association between online risks and mental health for youth. 

\subsection{Associations Between Online Risks and Mental Health (RQ2)}
\subsubsection{Investigating the Differences in Mental Health Outcomes based on Sexual Identity (H3).} 
Next, we investigated whether there were significant differences in mental health challenges reported by LGBTQ+ versus heterosexual youth. Our Welch’s Two-Sample \textit{t}-test compared between-group differences based on mental health scores and revealed a statistically significant difference in self-reported Depression, Self-Harm and Injury, and Mental Well-Being scores \textit{p} < .001, as illustrated in Table \ref{Mental Health Comparsion}. LGBTQ+ youth reported higher levels of depressive thoughts, Self-Harm, and Injury behavior than heterosexual youth. LGBTQ+ youth also scored significantly lower in Mental Well-Being than heterosexual youth. Therefore, the results confirmed our hypothesis (H3) that LGBTQ+ individuals report worsened mental health outcomes compared to heterosexual youth. These results serve to confirm past studies reporting similar results~\cite{dawson2023mental,utah2021healthcare}.

\begin{table*}[!h]
  \footnotesize
     \centering
  \caption{\textit{T}-test results comparing LGBTQ+ and heterosexual on mental health constructs}
  \label{Mental Health Comparsion}
  \begin{tabular}{lllllllll}
    \cmidrule{1-9}
    \textbf{Measure}&  \textbf{\textit{Cronbach's $\boldsymbol{\alpha}$}} &
    \textbf{Sexual Identity} & \textbf{\textit{Mean}} & \textbf{\textit{SD}} & \textbf{\textit{t}} & \textbf{\textit{df}} & \textbf{\textit{Std. CI}} & \textbf{\textit{p}}\\

    \midrule
    Depression & \edit{0.918}  & LGBTQ+ & 3.598 & 0.881 & -5.788 & 170.94 & -1.036 - -0.509 &  \textbf{<.001}\\
    & &Heterosexual & 2.825 & 0.876 \\
    \midrule
    Self-Harm and Injury & \edit{0.878} & LGBTQ+ & 1.858 & 0.773 & -4.908 & 142.47 & -0.676 - -0.287   & \textbf{<.001}\\
    & &Heterosexual & 1.375 & 0.484 \\
    \midrule
    Mental Well-Being & 0.897 & LGBTQ+ & 4.279 & 1.102 & 4.248 & 170.82 & 0.377 - 1.032  & \textbf{<.001}\\
    & & Heterosexual & 4.984 & 1.079 \\
    
  \bottomrule

\end{tabular}
\vspace{-4pt}
\end{table*}

\begin{table*}[htbp]
  \footnotesize
     \centering \caption{Unstandardized estimates for linear regression models examining the relationship between online risky messages (e.g., sexual messages, harassment, and self-injury) and mental health constructs (e.g., Self-Harm and Injury, Depression, and Mental Well-Being)}
  \label{sexual vs health}
 \resizebox{\textwidth}{!} {\begin{tabular}{lccccccccc}
    \toprule
    & \multicolumn{3}{c}{\textbf{Self-Harm and Injury (M1)}} & \multicolumn{3}{c}{\textbf{Depression (M2)}} & \multicolumn{3}{c}{\textbf{Mental Well-Being (M3)}} \\
    \textit{Predictors} &  \textit{Estimates} &  \textit{Std. CI} & \textit{p}  &  \textit{Estimates} &  \textit{Std. CI} &  \textit{p} &  \textit{Estimates} &  \textit{Std. CI} &  \textit{p} \\

    \midrule
    (Intercept) & 1.284 & 1.100 - 1.467 & <0.00*** & 2.639& 2.382 - 2.895 & <0.00*** & 5.175 & 4.858 - 5.492 & <0.00***\\
    \\
    \textbf{Sexual Messages} & \textbf{0.010} & \textbf{0.000 - 0.020} & \textbf{0.048\oneS} & \textbf{0.014} & \textbf{0.000 - 0.028} & \textbf{0.048\oneS} &  \textbf{-0.020} & \textbf{-0.037 - -0.002} & \textbf{0.026\oneS}\\
     \\
    Harassment & -0.000 & -0.008 - 0.007 &  0.865 & 0.002 & -0.009 - 0.013 & 0.717 & -0.001 & -0.015 - 0.012 & 0.803\\
    \\
    Self-Injury & -0.011 & -0.167 - 0.144 & 0.885 & 0.082 & -0.136 - 0.301 & 0.457 & 0.064 & -0.205 - 0.334 & 0.636\\
    \\
    \textbf{Sexual Identity (LGBTQ+)} & \textbf{0.328} & \textbf{0.052 -  0.605} & \textbf{0.020\oneS} & \textbf{0.891} & \textbf{0.504 -  1.278} & \textbf{< 0.001***} & \textbf{-0.994} & \textbf{-1.472 - -0.517} & \textbf{< 0.001***}\\
    \\
    Sexual Messages x Sexual Identity (LGBTQ+) & -0.006  & -0.019 - 0.007 & 0.379 & -0.012  & -0.031 - 0.006 & 0.200 & 0.028  & 0.004. - 0.051 & \textbf{0.018\oneS}\\
    \\
    Harassment x Sexual Identity (LGBTQ+)  & 0.020  & 0.003 - 0.037 & \textbf{0.018\oneS} & -0.003  & -0.020 - 0.027 & 0.749 & 0.001  & -0.028 - 0.031 & 0.919\\
    \\
    Self-injury x Sexual Identity (LGBTQ+)  & 0.058  & -0.108 - 0.225 & 0.489 & -0.088  & -0.322 - 0.145 & 0.458 & -0.063  & -0.352 - 0.225 & 0.663 \\
    \\
  \midrule
  Observations & \multicolumn{3}{c}{173} & \multicolumn{3}{c}{173} & \multicolumn{3}{c}{173}\\
  $R^2$ / Adjusted $R^2$ & \multicolumn{3}{c}{0.195 / 0.161} & \multicolumn{3}{c}{0.190 / 0.156} & \multicolumn{3}{c}{0.135 / 0.098}\\
      \bottomrule 
  \\
  \multicolumn{9}{l}{Note.  *\textit{p} < .05; **\textit{p}<.01; ***\textit{p}<.001}
  \\
\end{tabular}}
\end{table*}

\subsubsection{Examining the Associations between Online Risk and Mental Health Outcomes (H4).}
The results of the second chi-square test for RQ1 (\autoref{chi_square} (b)), which was conducted to examine the variation in risk types, guided our subsequent analyses for H4. Specifically, we focused our analysis on sexual messages, harassment, and self-injury-related online interactions due to the significant differences found between LGBTQ+ and heterosexual youth in relation to these distinct categories of risky online messages received. Therefore, we investigated to what extent the online risk experiences encountered by youth in private online spaces were associated with mental health outcomes (i.e., Self-Harm and Injury, Depression, and Mental Well-Being), and how this relationship was moderated by participants' sexual identity.

\textbf{\emph{Self-Harm and Injury:}} Our first multiple regression model (M1) identified a significant and positive association between messages containing sexual content and youths' Self-Harm and Injury scores (see Table \ref{sexual vs health}). Results indicated that youths more frequently engaged in self-harming behaviors when they also received more sexual messages through private messages, regardless of sexual identity (\textit{t} = 1.991, \textit{df} = 165, \textit{p} = .048, 95\% confidence interval: [0.000, 0.020]). In contrast, harassing messages and those having to do with self-injury did not predict Self-Harm and Injury mental health scores (\textit{p} > .05). \edit{Additionally, we conducted a detailed examination of individual Self-Harm and Injury behaviors by implementing a multivariate regression model (M4) to discern the specific self-harm behaviors that significantly impact youth's mental health. The outcomes of our M4 model align with those of the M1 model, reinforcing earlier findings. Notably, we observed a positive and statistically significant main effect between messages containing sexual content and particular self-harm behaviors, such as carving and pulling hair, aligning with M1. Additionally, the model (M4) reveals a significant interaction effect between messages containing harassment and behaviors such as cutting, carving, and rubbing skin against a rough surface, specifically for LGBTQ+ youth. However, no significant effects were identified for other risky messages and their association with Self-Harm and Injury behaviors (p > .05). \hyperref[appendix:a]{Appendix A} provides comprehensive details of this model.}
\vspace{5pt}

\textbf{\emph{Depression:}} In our second regression model (M2), a significant correlation emerged between messages featuring sexual content and the self-reported Depression scores of young individuals (see Table \ref{sexual vs health}). The positive direction suggested that regardless of their sexual identity, young individuals reported more frequent experiences of depressive thoughts when they received a higher volume of sexual messages through private messaging (\textit{t} = 1.992, \textit{df} = 165, \textit{p} = .048, 95\% confidence interval: [0.000, 0.028]). Conversely, messages that included harassment or language likely to induce self-injury did not forecast self-reported Depression scores in young individuals (\textit{p} > .05).

\textbf{\emph{Mental Well-Being:}} In our third regression model (M3), we observed a significant negative association between messages containing sexual content and the Mental Well-Being scores of young individuals (see Table \ref{sexual vs health}). The findings revealed that irrespective of their sexual identity, young individuals reported lower scores, indicating a decline in Mental Well-Being, when they received an increased number of sexual messages via private messaging (\textit{t} = -2.244, \textit{df} = 165, \textit{p} = .026, 95\% confidence interval: [-0.037, -0.002]). On the other hand, messages that contained harassment or language that might prompt self-injury did not predict the Mental Well-Being scores of young individuals (\textit{p} > .05). \edit{According to established standards in social science research an $R^2$ value of 10\% or above is considered acceptable~\cite{ozili2023acceptable, falk1992primer}. Our models, across all three mental health measures (Mental Well-being marginally close to the acceptable threshold), meet this threshold, signifying that they provide meaningful insights into the relationships between the predictors and mental health outcomes.}

\edit{In summary, our results (M1, M2) reveal that a substantial correlation exists between youth encountering a significant number of messages containing sexual content and reporting more pronounced mental health challenges (e.g., higher Depression scores, increased Self-Harm and Injury), and lower Mental Well-Being scores (M3). This association stands out, as other types of risky encounters, such as messages containing harassment and self-injury, did not demonstrate statistical significance with any of the mental health constructs for the youth as a whole. Consequently, our multiple linear regression models provide insightful findings, partially confirming our hypothesis (H4), emphasizing that youth with more high-risk online experiences tend to report more adverse mental health outcomes compared to those with lower levels of online risk. Next, we report on the moderating effects of our models.}

\subsubsection{Examining the Moderating Effect of Online Risk and Sexual Identity on Mental Health (H5).}
\begin{figure*}
\footnotesize
    \centering
\begin{subfigure}[t]{.45\textwidth}\centering
  \includegraphics[width=\columnwidth]{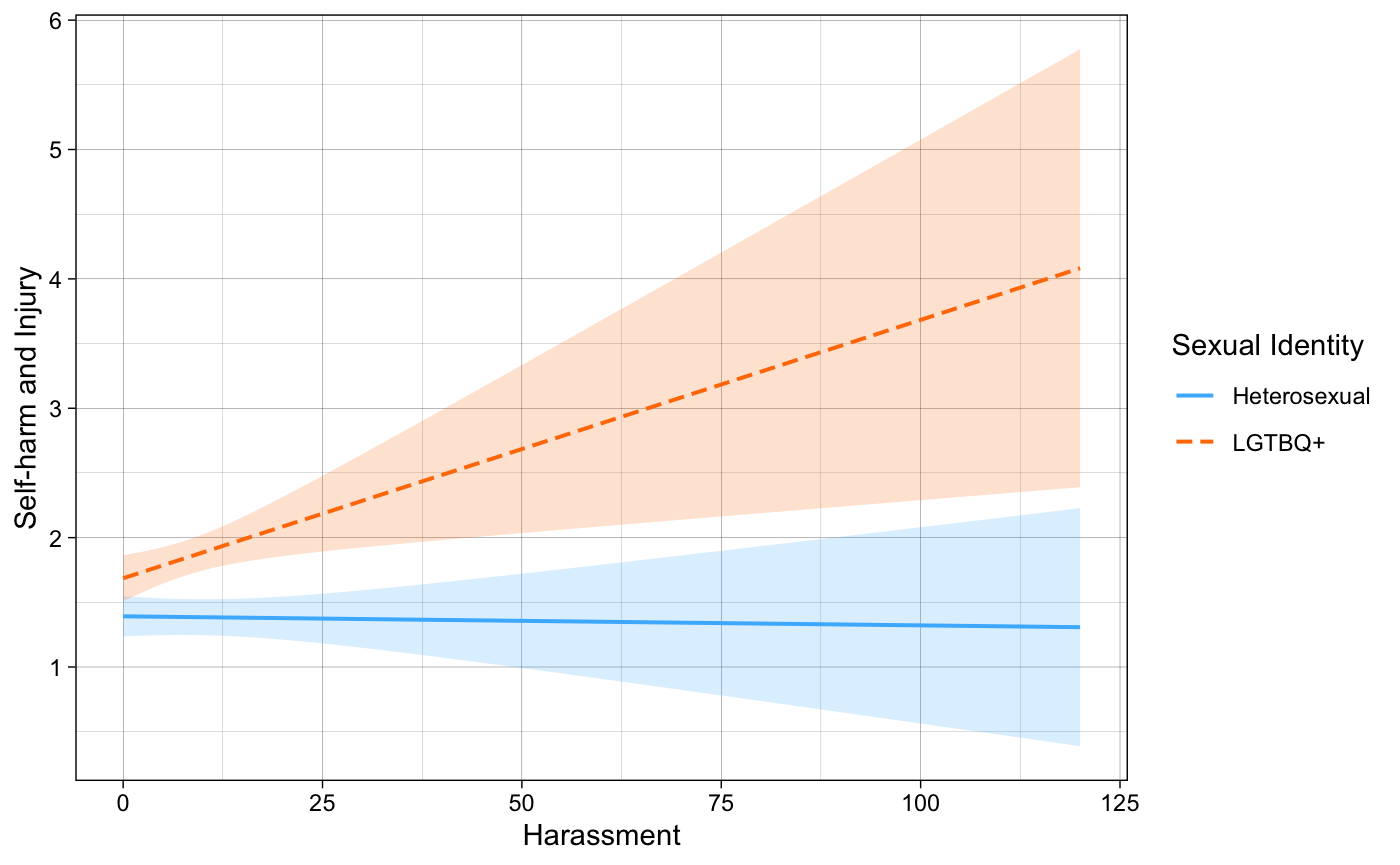}
    \caption{}
\end{subfigure}%
\hspace{1em}
\begin{subfigure}[t]{.45\textwidth}\centering
   \includegraphics[width=\columnwidth]{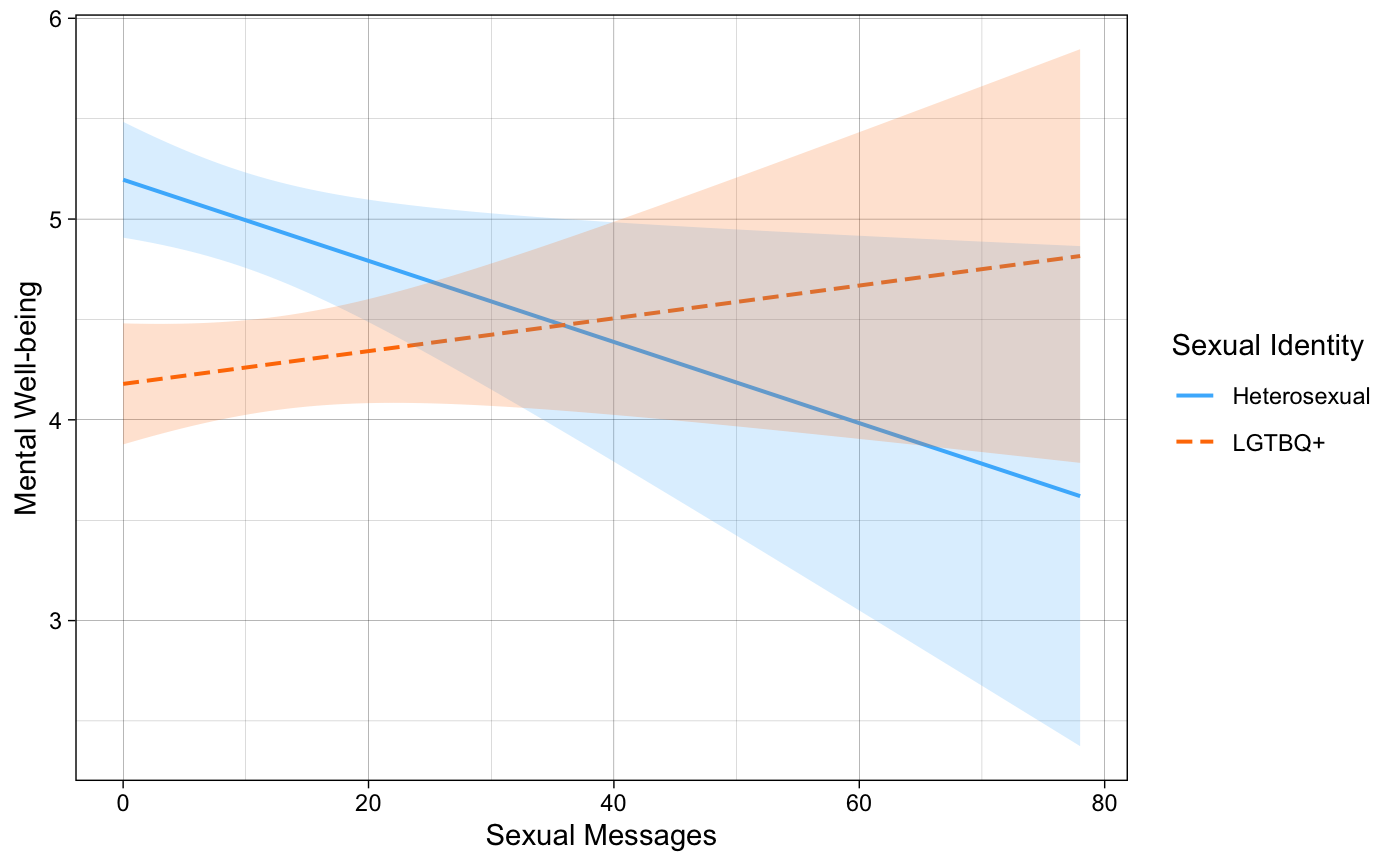}
   \caption{}
 \end{subfigure}
\centering 
    \caption{Moderating Effect of self-reported sexual identity on the association between online risk experiences and mental health outcomes: (a) Shows a positive relationship between harassment (X-axis) and Self-Harm/Injury (Y-axis); (b) Demonstrates a positive relationship between sexual messages (X-axis) and Mental Well-being (Y-axis) for LGBTQ+ youth. Orange and blue lines depict the moderating effect for LGBTQ+ and heterosexual youth, respectively.}~\label{interaction_1}
\end{figure*}

We examined the moderating effect of sexual identity on the relationship between risky online messages and youth's mental health outcomes. In doing so, our multiple linear regression model (M1) identified a significant interaction effect between harassing messages and Self-Harm and Injury such that the non-significant main effect held for heterosexual youth; yet, the correlation between harassing messages and Self-Harm and Injury for LGBTQ+ youth became significant and positive. Figure \ref{interaction_1} (a) shows the significant interaction effect between harassment messages and Self-Harm and Injury behavior of youth. Results indicated that harassment within direct messages represents a unique risk factor for Self-Harm among LGBTQ+ youth (\textit{t} = 2.386, \textit{df} = 165, \textit{p} = .018, 95\% confidence interval: [0.003, 0.037]). In other words, harassment is associated with an increase in Self-Harm for young people who identify as LGBTQ+, whereas heterosexual youth are less affected. Hence, this finding supports our hypothesis (H5), which is the relationship between online risks (e.g., harassment) and adverse mental health outcome (e.g., Self-Harm and Injury behavior in this instance) will be stronger for LGBTQ+ youth. In contrast, the non-significant interaction effect between sexual messages and sexual identity indicates that sexual messages are positively associated with Self-Harm and Injury behaviors for LGBTQ+ and heterosexual youth alike. Similarly, there was no significant interaction effect found between messages containing self-injury language and sexual identity in relation to Self-Harm and Injury behaviors (\textit{p} > .05).

For our model (M2) with Depression score as the outcome variable, the interaction effect between risky messages (e.g., sexual messages, harassment, and self-injury-containing messages) and Depression scores was not statistically significant (\textit{p} > .05). The lack of an interaction effect between sexual messages and sexual identity suggests that sexual messages can be concerning for both LGBTQ+ and heterosexual youths when it comes to their depressive thoughts. 

On the other hand, our model (M3) identified a significant relationship between messages containing sexual content toward LGBTQ+ youth and Mental Well-Being scores. \edit{Figure \ref{interaction_1} (b) shows the significant interaction effect between sexual messages (X-axis) and the Mental Well-Being score (Y-axis) of youths. The results indicated that sexual messages are uniquely associated with positive Mental Well-Being for LGBTQ+ youth only (\textit{t} = 2.386, \textit{df} = 165, \textit{p} = .018, 95\% confidence interval: [0.004, 0.051]). In other words, sexual messages are associated with a decrease in Mental Well-Being scores for heterosexual young people (blue straight line), while for LGBTQ+ youth (orange dotted line) the association is in the opposite direction.} Consequently, this finding contradicts our hypothesis (H5), which stated that the relationship between online risks (e.g., sexual messages) and negative mental health outcomes (e.g., Mental Well-Being scores) would be more pronounced for LGBTQ+ youth compared to heterosexual youth. However, there was no significant interaction effect when examining the relationship between messages containing harassment or self-injury inducing language and sexual identity in relation to youth's self-reported Mental Well-Being scores (\textit{p} > .05). 

\edit{In summary, our investigation into how sexual identity moderates the relationship between negative mental health outcomes and online risk experiences yields novel insights. Our findings illuminate the distinctive role of harassment as a risk factor for Self-Harm and Injury, while also uncovering an unexpected positive association between sexual messages and the Mental Well-Being of LGBTQ+ youth. Therefore, our results contribute to a nuanced understanding and partially support our hypothesis (H5) that the link between online risks and adverse mental health outcomes will be more pronounced for LGBTQ+ youth. Table \ref{hypotheses} provides a summary of the hypothesis testing results from our statistical model.}

\begin{table*}[!h]
  \footnotesize
     \centering
  \caption{Hypotheses testing results}
  \label{hypotheses}
  \begin{tabular}{lclc}
    \toprule
    \textbf{Hypotheses} & \textbf{Supported}\\
    \midrule
    H1: LGBTQ+ youth will report more online risks than heterosexual youth & No \\
    H2: LGBTQ+ youth will report more high risk online experiences than heterosexual youth & Yes \\
    H3: LGBTQ+ youth will report worse mental health outcomes than heterosexual youth & Yes \\
    H4: Youth with more high risk online experiences will report worse mental health outcomes & Partially  \\
    H5: LGBTQ+ youth will exhibit a strong relationship between online risks and negative mental health outcomes & Partially \\
  \bottomrule
\end{tabular}
\vspace{-3mm}
\end{table*}

\subsection{Nature of Risks Experienced by LGBTQ+ Youth in Private Online Spaces (RQ3)}

To further unpack the statistically significant effects found in RQ1 and RQ2, this section qualitatively examines risk-flagged Instagram conversations donated by the LGBTQ+ youth to shed light on the nature of online risks. LGBTQ+ youth participants flagged a total of 1,468 private messages of 808 conversations as "risky", and these conversations serve as the unit of analysis for the results presented below. Table \ref{tab:codebook} presents the themes and subthemes for RQ3, as well as their corresponding codes and illustrative quotations. We used illustrative quotations to describe each of the themes that emerged from our thematic analysis. In the illustrative conversations, we replaced all referenced names with the letter "X" in the messages and removed the Unicode characters (emojis). Overall, we found that the risky experiences reported by LGBTQ+ youth were often related to their sexual identity.

\subsubsection{Most Risky Conversations had Sexual Undertones.}
Overall, we found that most of the conversations LGBTQ+ youth flagged as making them feel unsafe or uncomfortable contained sexual content, either related to their gender or sexual identity and/or made direct sexual solicitations. For instance, more than one-third of the conversations contained \textit{messages with sexual content},  and one-fourth of these conversations took place \textit{with strangers}. Sexual messages often contained explicit adult content (e.g., photos and videos), porn website URLs, and sexual texts. For example, LGBTQ+ youth were often added to porn group chats by strangers, where they also received inappropriate content and/or porn site URLs. Interestingly, we often found LGTBQ+ youth were interested in participating in such conversations as a way to explore their sexual identities but wanted to do so slowly. For example, a 21-year-old homosexual woman had the following conversation with a stranger:

\begin{quote}
\textbf{Other person:} \textit{You are so incredibly beautiful. Its an absolute pleasure to meet you. What exactly is it you are looking for? I'm looking for a special friend to share, maybe explore things together. We can be friends and forward funny videos. You feel you might be interested in something casual like this?} \newline
\textbf{Participant:} \textit{Uh sure, Haha, Omg! I think I probably seen you on a dating app lol} \newline
\textbf{Other person:} \textit{Instagram User sent an attachment} [flagged as sexual content] \newline
\textbf{Participant:} \textit{whoa, thats just downright fast}

\end{quote}

While the LGBTQ+ youth mostly received sexual message content from strangers, we found some conversations where participants \textit{received sexual messages from their peers} and the \textit{youth themselves also sent sexual messages to others}. When asked to describe the situation, they explained that they sent these sexual photos either to experiment or to earn money, but often later they regretted it. Below is an example conversation that a 21-year-old bisexual woman flagged: 

\begin{quote}

\textbf{Other Person:} \textit{Do u have haire on pussy } \newline
\textbf{Participant:} \textit{A lil } \newline
\textbf{Other Person:} \textit{Can I see } \newline
\textbf{Participant:} \textit{X sent an attachment} [flagged as sexual content] \newline
\textbf{Other Person:} \textit{and boobs}
\end{quote}

Almost a quarter of the conversations consisted of text messages with \textit{sexual solicitations}. These conversations frequently included adult content; therefore, they were double-coded for sexual content and solicitations. Similar to the sexual message content, LGBTQ+ youth mostly received these \textit{sexual solicitations from strangers}. In some cases, participants were offered different kinds of incentives, e.g., \textit{monetary and/or other benefits}, in exchange for a sexual relationship with these strangers. For instance, a 19-year-old bisexual woman received the following messages:
\begin{quote}
\textbf{Other Person:} \textit{Hello beautiful, can you be my sugar baby? I'm ready to help you and you gonna get paid weekly. Let me know when you're ready. Just chat with me everyday and you get your paid for that} [flagged as harassment] \newline
\textbf{Participant:} \textit{can u buy my amazon wish list} \newline
\textbf{Other Person:} \textit{Money has never been my problem,How much is your amazon wishlist} \newline
\textbf{Participant:} \textit{like \$300ish}
\end{quote}

Similar to the sexual content, LGBTQ+ youth did not just receive these sexual solicitations from strangers, but some of the sexual solicitations came \textit{from their peers} as well. While in majority of the conversations we saw youth to participate or interact with the other person, there were a few other conversations where we found our participants \textit{received persistent unwanted messages} that they flagged as "harassment". These messages were received from strangers and peers both. Although youth repeatedly requested to stop texting, they kept sending these messages. Interestingly, we noticed there were a small portion of conversations where they started with sexual content and/or sexual solicitations, but later they \textit{moved toward harassment}. This is because participants often refused to advance with the sexual solicitations, which caused the other person to become aggressive, sometimes even threatening the youth.

Along with the above sexual solicitations, we found conversations where participants were \textit{harassed because of their sexual identity}. Participants often labeled these messages as "harassment" or "unwanted messages". They mostly received these messages from their peers. Some of these conversations also had sexual slurs mentioned within instances of \textit{sexual harassment}. For example, a 20-year-old transgendered man flagged the following conversation that they had with someone they knew:

\begin{quote}
\textbf{Other Person:}	\textit{Call me.} [Video call started.] [Video call ended.]\newline
\textbf{Participant:}	\textit{Wtf what’s wrong with you} \newline
\textbf{Other Person:}	\textit{I didn’t mean to do that} \newline
\textbf{Participant:}	\textit{Oh really?} [Video call started.] [Video call ended.]\newline 
\textbf{Other Person:} \textit{Answer the phone fag} [flagged as harassment] \newline
\textbf{Participant:}	\textit{Stop texting me please}
\end{quote}

LGBTQ+ youth also received similar messages from strangers where they were harassed for their sexual identities. Interestingly, most of these conversations took place in large group chats created for makeups, games, travels, electronics troubleshooting, etc, where people may not know each other. Youth were often harassed during casual conversations related to technical problems with their games and/or computers. For instance, a 21-year-old gender-fluid woman had the following conversation in a group chat for gamers: 

\begin{quote}
\textbf{Participant:}	\textit{My fucking laptop is just burning up} \newline
\textbf{Other Person:}	\textit{thats because you're a faggot and your gonna burn in hell} [flagged as harassment]\newline
\textbf{Participant:}	\textit{well that was impolite because you dont know anythin abt me} 

\end{quote}

LGBTQ+ youth also often \textit{received different advertisements to promote sexual products and services that were specifically targeted to their sexual identities}. Participants mostly flagged these messages as spam or harassment. In these cases, it seemed apparent that these advertisements and/or scams were aware of their sexual identities, possibly disclosed through their profile descriptions or public posts. LGBTQ+ youth also \textit{received scam messages} that often contained a monetary offer for signing up on a porn website, sending body photos, or working as a sex worker. Participants mostly did not reply to these messages; therefore, we did not find any conversations where they stepped into these monetary traps. In fact, many of the youth seemed savvy to avoid or exploit these solicitations to their own advantage. Below is an example conversation that a 21-year-old transgender (bisexual) woman had with a stranger:
\begin{quote}
\textbf{Other Person: } \textit{Hi,  My ex trans baby left me. Will you be my sugar baby } [flagged as spam] \newline
\textbf{Participant: } \textit{pay me first ? bc i’ve been in these situations b4} \newline
\textbf{Other Person: } \textit{ I’m a legit sugar daddy} \newline
\textbf{Participant: } \textit{ send me \$100 den}
\end{quote}

As shown above, LGBTQ+ youth often received sexual content and solicitations from strangers, but they often took these interactions in stride until they became threatening. However, when they were harassed because of their sexual identities, they did not appreciate such messages. In the next section, we present other general types of online risks that LGBTQ+ youth often reported. 

\subsubsection{LGBTQ+ Youth Encountered Harassment and Scams.}
LGBTQ+ youth also received other types of risky messages that were not sexual in nature. For example, we found one-third of the conversations had some form of harassment, e.g., namecalling, threats, abusive words. Participants received these messages not just from strangers but from their peers also. We noticed that youths were often bullied (with threats and abusive words) by peers because of arguments over some physical incidents that happened at the their schools or workplaces. Also, they were often body-shamed by their peers. For instance, a 15 years old non-binary (bisexual) woman had the following conversation with a peer:

\begin{quote}
\textbf{Other Person:} \textit{hey, your kind fat. you should lose weight. I remember you lost weight... what happened} [flagged as harassment] \newline
\textbf{Participant:} \textit{um it was hard for me. I've been suffering... Stop sending me hate messages}
\end{quote}

Besides the harassing messages, participants also received \textit{different advertisements} and \textit{scam messages} that were not specifically targeted to their sexual identity. LGBTQ+ youth flagged these unsolicited messages as unsafe because they often promoted different products or websites and/or intended to cause financial harm. Some of these spam messages contained links to the websites of \textit{illegal products}, e.g., tobacco or recreational drugs. Participants rarely responded to these messages, and therefore, these conversations were mostly one-way. Because these risk experiences seemed relatively typical to the risks generally experienced by youth online \cite{10.1145/3491101.3503569, 10.1145/3479609, 10.1145/3449116, 10.1145/3334480.3383073}, we chose not to examine these conversations in more depth. 

\subsubsection{LGBTQ+ Youth Received Support from Peers Regarding Negative Experiences.}
Although our conversation dataset focused on risky messages flagged by the LGBTQ+ youth, we often found messages in these conversations where the LGBTQ+ youth \textit{shared negative experiences relating to their sexual identities with peers}. Most of these conversations had messages that they flagged as harassment or bullying. This is because, in the same messaging thread, the youth had arguments with their peers over an issue that led to harassment, and therefore, they flagged the messages as unsafe. However, the youth also shared their negative emotions with the same peers, receiving compassion and moral support. Participants were also seen to share \textit{experiences that they had with homophobic people around them}, mostly in their families. In some of these conversations, they expressed that they are \textit{afraid of coming out} because of their LGBTQ+ intolerant families. Participants also often discussed their \textit{urge to self-harm} as they became depressed about their daily struggle for their gender identities. Interestingly, youth often mentioned about the negative experiences they faced when they publicly shared their photos on different social media platforms. Below is an example conversation that a 19-year-old gender-fluid man had with a peer:

\begin{quote}
\textbf{Participant:}\textit{ i wish things were going ok i have been having thoughts of wanting to cut and i freaking hate it so much i just want to scream and cry for hours}\newline
\textbf{Other Person:}	\textit{I’m sorry to hear it, self-harm urges are awful} \newline
\textbf{Participant:}\textit{ I hate that i am still afraid to posting wearing a dress or skirt it is making me dysphoric as fuck. People are good for judging me for showing my pics on Instagram and This is the 5th time I was treated badly online And im so fucking tired of it. }
\end{quote}

Participants did not just share their struggles related to gender identity; they also often discussed their mental health issues with peers. LGBTQ+ youth expressed their sad feelings about different negative incidents that occurred with their friends and family. Besides these, participants also \textit{shared their depressive thoughts}, e.g., self-harm, suicide, with their peers. In all these messages, we saw their peers being compassionate and providing advice to cope. For instance, a 15-year-old homosexual woman and their peer had the following conversations:

\begin{quote}
 \textbf{Participant: }\textit{ And maybe I’d been lucky to die} \newline
\textbf{Other Person:} \textit{Umm X, don’t die on me}\newline
 \textbf{Participant: }\textit{ You know almost everytime I smile or something of that sort is fake, because I’ve been broken for so long I’ve become the person I thought I was when I was 9-12. I’ve become the girl who literally lost it from grief, and wants to just end it all} \newline
\textbf{Other Person:} \textit{Wait X plz no crying...imma bout to be sad now}
\end{quote}

Alongside the depressive feelings, we also often found the LGBTQ+ youth sharing other feelings about the \textit{negative experiences} that they had with their families, friends, and others. We observed that these feelings were not sad in particular, but more related to fear, anger, disgust, and/or anxiety. Participants also occasionally \textit{exchanged advice on their drug usage}. A few conversations contained messages where one person used illegal tobacco or drugs and the other person gave advice or showed concern for their risky behavior. In summary, LGBTQ+ youth did not just receive risky messages from others; they also received support and advice from their peers about their negative experiences with their family and friends. More importantly, they often sought support for the struggles that they dealt with in their families regarding their sexual identity.

\section{Discussion}

In this section, we briefly summarize our key findings, unpack the implications of these results, and provide recommendations for empowering and protecting LGBTQ+ youth in social media spaces.

\subsection{LGBTQ+ Young Experience More Severe Online Risks on Social Media (RQ1)}

While LGBTQ+ youth in our study flagged fewer Instagram DMs as risky (48.56\%) compared to heterosexual participants (H1), they encountered significantly more high-risk situations (H2), involving increased instances of sexual messages and self-injury. Conversely, heterosexual youth reported more low-risk conversations, primarily centered around online harassment. Despite conflicting with previous research that positioned LGBTQ+ youth as highly vulnerable to cyberbullying ~\cite{aboujaoude2015cyberbullying, zych2015systematic, zych2016cyberbullying}, our qualitative findings elucidate this apparent contradiction. Many instances of online harassment experienced by LGBTQ+ participants were sexually motivated, leading to categorization as sexual rather than general harassment. This insight reveals the unique challenges faced by LGBTQ+ youth, who often endure online harassment targeting their sexual and gender identities, arguably more detrimental than bullying experienced by heterosexual counterparts. Consequently, our study indicates that LGBTQ+ youth reported significantly more risky conversations involving self-injury and mental health struggles compared to their heterosexual counterparts.

These significant findings underscore the need for urgent action in implementing targeted risk prevention and mitigation programs addressing sexually motivated online harassment and supporting the mental health of LGBTQ+ youth both online and offline. Existing national programs like "Netsmartz" in the United States ~\cite{netsmartz2023} and "ThinkUKnow" in the United Kingdom ~\cite{thinkuknow2023} offer commendable educational resources but primarily target broader audiences. To effectively safeguard vulnerable communities, such as LGBTQ+ youth, there's a growing demand for specialized education programs tailored to their unique risks and challenges. While organizations like The Trevor Project provide vital crisis intervention and suicide prevention services for LGBTQ+ youth ~\cite{trevor1998}, recent studies highlight limitations, including resource availability and operational challenges~\cite{levtove2023}. Establishing such platforms alone is insufficient; continuous monitoring and timely support are crucial for their effectiveness. Our research emphasizes the need for more focused studies on vulnerable youth, including LGBTQ+ ~\cite{gatehouse2018troubling,saha2019language}, neurodiverse ~\cite{page2022perceiving,barros2023my}, and foster youth ~\cite{badillo2019risk,fowler2022fostering}, concerning their online safety. Recent discourse on youth empowerment and resilience in online spaces~\cite{agha2023strike, akter_from_2022, wisniewski2015resilience,  akter_it_2023, akter_evaluating_2023, kumar2022tiktok, akter_CO-oPS_2022} may not be inclusive of the needs and struggles of more vulnerable youth populations, necessitating a more comprehensive approach.

\subsection{Social Media's Dual Impact on LGBTQ+ Youth Mental Health (RQ2)}

Our H3 results confirmed existing research indicating that depression disproportionately affects LGBTQ+ minority youths compared to their heterosexual counterparts ~\cite{quarshie2020prevalence}. Furthermore, LGBTQ+ youth exhibit higher rates of self-harm and diminished mental well-being ~\cite{almeida2009emotional,charmaraman2021young} when compared to heterosexual peers ~\cite{russell2016mental,russell2019sexual} (refer to Table \ref{Mental Health Comparsion} showing the differences in mental health constructs). In addition, our H4 and H5 results introduced novel insights by linking online risk experiences to mental health outcomes for both LGBTQ+ and heterosexual youth, addressing a crucial gap in the literature. Through our models M1-M3 (refer to Table \ref{sexual vs health}), we observed that sexual messages are significantly and positively associated with increased levels of self-harm, injury, and depression, while negatively impacting mental well-being for all youth, regardless of sexual identity. This supports our H4 that more high-risk online experiences correlate with worse mental health outcomes. However, examining interaction effects (H5) revealed nuances. Online harassment was significantly and positively linked to self-harm and injury (Figure \ref{interaction_1}) for LGBTQ+ youth, not evident in the main effects, while an opposite effect emerged for sexual messages: more messages were associated with reduced mental well-being for heterosexual youth but positively correlated with mental well-being for LGBTQ+ youth.

Careful consideration of these findings is essential due to the conflicting mental health outcomes associated with receiving sexual messages. Both LGBTQ+ and heterosexual youth showed higher rates of Depression and Self-Harm/Injury in association with sexual messages, yet for LGBTQ+ youth, an increase in sexual messages was also linked to improved Mental Well-Being. Our exploration of qualitative insights in response to RQ3 offers a potential explanation for this paradox. While LGBTQ+ youth faced sexual messages harassing their identities, they also received messages facilitating exploration without judgment. Future studies should distinguish between \textit{harmful} and \textit{helpful} sexual messages sent to and by LGBTQ+ youth. This holds crucial implications for Human-Centered Machine Learning (HCML) researchers developing algorithms to detect and mitigate online sexual risks ~\cite{razi2020let,razi2021human,razi2023sliding}. If these algorithms inadvertently restrict or censor beneficial sexual conversations of LGBTQ+ youth, the technologies designed to protect them may disproportionately harm them. Our research underscores the need to provide LGBTQ+ youth a safe space for exploring their sexual and gender identities. Future HCML research should strive to differentiate between sexually motivated harassment and violence online and healthy sexual exploration during adolescent developmental growth.

\subsection{Social Media Amplifies Risks But Provides Needed Support for LGBTQ+ Youth (RQ3)} 
Our qualitative findings played a crucial role in elucidating unexpected and occasionally conflicting results from our statistical analyses, offering deeper insights into private online interactions that made LGBTQ+ youth feel uncomfortable or unsafe. One key theme revealed in risk-flagged messages was the prevalence of sexual undertones, even in harassing messages, intensifying the interconnectedness between online risk and the sexual identities of LGBTQ+ youth. This underscores the need for spaces where LGBTQ+ youth can interact online without their sexual identities being the focal point, a benefit often naturally afforded to heterosexual youth. Our results emphasize the urgent societal need for increased awareness and education about the LGBTQ+ community to foster acceptance and inclusion. Additionally, our research highlights the normalcy of some level of risk-seeking behavior in youth development \cite{baumrind_developmental_1987}, as evidenced by LGBTQ+ youth's willingness to engage with flagged sexual messages, sometimes benefiting from these interactions. However, a concern arises as these conversations often occur with strangers, posing potential risks, including sexual predation. To address this, it is crucial to provide safe outlets for LGBTQ+ youth~\cite{betterhelp2022}, allowing them to question, explore, and discuss their sexual identities online.

Our research also revealed that LGBTQ+ youth flagged messages as risky even when sharing struggles with peers to seek support, highlighting the double-edged nature of social media. This aligns with existing literature indicating that LGBTQ+ individuals often utilize social media as a support group to share unique experiences both online and offline~\cite{craig2014you, lucero2017safe}. Notably, a proposed Kids Online Safety Act (KOSA)~\cite{Blumenthal2022} aims to enhance minors' online safety but raises privacy concerns due to increased surveillance and potential content filtering. Legislative efforts to ban certain platforms could inadvertently harm marginalized groups, such as LGBTQ+ teens who rely on social media to connect to peers ~\cite{Kelly_tennessee_2023, brown_clarksville_2023}, to question, explore, and understand their sexual identities ~\cite{Feiner2022}. Restricting Internet use~\cite{reed_senator_2023} for LGBTQ+ youth might diminish their support networks, necessitating practical solutions that reduce harmful risk exposure while providing opportunities for online support.

\subsection{Implications for Practice and Design}

We provide several actionable recommendations towards education and design in promoting the online safety and digital well-being of LGBTQ+ youth.

\subsubsection{Establishing Stronger Community Guidelines and Norms to Protect LGBTQ+ Youth Online.} Recent research highlights significant safety risks for LGBTQ+ users on social media platforms ~\cite{silvia2021}, with claims that these platforms prioritize profit over LGBTQ+ safety and lives ~\cite{yurcaba2022}. Moreover, researchers have raised doubts about the effectiveness of existing community guidelines ~\cite{connellan2022social,yurcaba2022}, emphasizing their inadequacy in protecting the LGBTQ+ community. Our study advocates for stronger community-based guidelines, specifically focusing on safeguarding sexual minority social media users, especially LGBTQ+ youth. These guidelines provide a crucial foundation for curtailing harmful content targeting sexual minority youths, setting clear boundaries for acceptable behavior, and discouraging harassment and discrimination. They convey a powerful message affirming the right to a safe online environment for all users, irrespective of sexual identity. Additionally, guidelines offer a structured mechanism for reporting unacceptable behavior, empowering victims to seek assistance. Violators can face consequences, including warnings, suspensions, or bans from the platform ~\cite{reddit2023content}. In summary, when combined with effective enforcement, these guidelines contribute to positive community norms by setting clear expectations, reinforcing social norms, fostering trust, and promoting responsible behavior ~\cite{scharlach2023governing, reddit2023coc}.

\subsubsection{Providing Resources for Cyberabuse Prevention and Support.}
To this end, providing help resources and establishing peer support networks for LGBTQ+ youth who are victims of cyberabuse is crucial. These resources empower youth to combat the effects of online abuse and seek help when needed. Equipping them with tools to address cyberabuse is paramount in giving them control over their online experiences. Previous research delved into the effectiveness of diverse support systems, encompassing family, curriculum, peer networks, school policies, Gay-Straight Alliances, etc. and confirmed that these elements are positively linked to the enhancement of positive socioemotional, behavioral, and educational outcomes for LGBTQ+ youth~\cite{leung2022social}. Additionally, we must establish mechanisms for bystander intervention to create a safer online environment. Encouraging friends, family members, and online community members to step in and advocate for LGBTQ+ youth ensures that they do not bear the burden of responsibility alone. In sum, our research serves as a clarion for immediate action, emphasizing the necessity of creating safer and more supportive digital spaces for all youth, with particular attention to the unique needs and challenges faced by LGBTQ+ individuals.

\subsubsection{Creating Safe Online Spaces for LGBTQ+ Youth.} 
Our research underscores the need for safe online spaces explicitly designed for LGBTQ+ youth. These spaces should serve as shelter, where they can explore their sexual identities and access accurate sexual health information without fear of judgment. The absence of such secure environments often turn LGBTQ+ youth to seek information and support from fringe or unregulated online communities~\cite{badham2021lgbtq,drug2022fight}. Unfortunately, these spaces may lack adequate moderation, exposing youth to harmful ideologies and potential safety risks. Furthermore, our findings highlight the significant mental health challenges faced by LGBTQ+ youth, who reported a higher prevalence of such challenges compared to heterosexual youth. This highlights the urgency of ensuring easy access to professional support resources in online settings to mitigate these adverse health outcomes. Such initiatives can act as protective measures, reducing the need for LGBTQ+ youth to engage with potentially harmful strangers online. In essence, our research serves as a call to action to create safer and more supportive digital environments for all youth, with a particular focus on the unique needs and challenges faced by LGBTQ+ individuals~\cite{trevor1998,lgbt2023,trans2023}.

\subsubsection{Developing Automated Risk Detection Differentiating between Sexual Harassment and Exploration.}
In contrast to our previous point, we also advocate for existing social media spaces to be safer for LGBTQ+ youth, so that they do not need to segregate themselves. Our findings shed light on the heightened vulnerability of LGBTQ+ youth to unsolicited \edit{sexual} messages from strangers. Consequently, there is a pressing need for the development and implementation of automated risk detection systems capable of identifying cyberabuse directed at the LGBTQ+ community and enforcing appropriate penalties. \edit{Navigating this challenging landscape reveals algorithmic bias as a significant threat to the well-being of LGBTQ+ individuals~\cite{let2022}. Past studies have shown that automated content moderation often restricts LGBTQ+ content, under the guise of "preserving decency" and "protecting the youth"~\cite{AI2021,automated2020,let2022}. The challenges associated with bias detection involve anomaly identification and assessing error rate equality~\cite{lee2019algorithmic,zafar2017fairness,spielkamp2017,davies2016}. HCML emerges as a pivotal solution, emphasizing fair language models and comprehensive training in gender-neutral pronouns~\cite{chancellor2023toward,tomasev2021fairness}. HCML strategies include expanding nondiscrimination laws, regulatory sandboxes, safe harbors, and self-regulatory practices with bias impact statements and inclusive design. Efforts also focus on algorithmic literacy and collaborative mitigation through formal feedback mechanisms~\cite{lee2019algorithmic}.} However, a significant challenge lies in ensuring that these penalties do not inadvertently infringe upon or harm the LGBTQ+ population. Striking a delicate balance is essential to mitigate potentially harmful interactions while preserving the rights of LGBTQ+ individuals to freely express their sexual identity. These automated systems should demonstrate sophistication in distinguishing between different types of content. For instance, they should distinguish between hate speech directed towards these communities and content that addresses transgender or homosexual issues.

\subsubsection{Raising Awareness and Inclusivity.} 
Our study underscores the need to foster greater acceptance and understanding of the LGBTQ+ community. Achieving this goal requires a concerted effort to expand online awareness and educational programs. A diverse array of strategies may be employed to cultivate more inclusive and empathetic environments for LGBTQ+ youth. One strategy may involve educational campaigns for disseminating accurate and up-to-date information regarding LGBTQ+ issues. Previous research highlights how pervasive myths can undermine and dehumanize LGBTQ+ individuals, often unfairly labeling these young individuals as ``confused'' or ``misguided''~\cite{abreu2017myths}. Therefore, we suggest that these campaigns adopt a conservative approach that incorporates storytelling and representation supported by facts from evidence-based research. Such personal narratives, combined with objective research, have the potential to put faces and voices to humanize LGBTQ+ experiences, making it easier for others to empathize and relate~\cite{suzuki2018dialogues,bauer2003power}. A social media campaign explaining `deadnaming' and how it is harmful to transgender people would be a specific example~\cite{resnick2023deadnaming}. This approach has the potential to address common misconceptions, ultimately enlightening the public on the complexities of LGBTQ+ identities.

\subsection{Limitations and Future Research}
We would like to highlight some limitations of our work, which inform future research directions. First, since youth participants were asked to flag at least two conversations that made them feel unsafe or uncomfortable, it is possible that they did not flag all conversations that met this criteria. Further, because we allowed youth to self-annotate their risk flagged messages by risk level and type, we do not account for individual differences in their perceptions of risk. At the same time, we adopt a victim-centered lens that believes participants' lived experiences, rather than questioning these experiences. Second, to have enough power to detect significant differences between heterosexual youth and LGBTQ+ youth, we had to group LGBTQ+ youth allies (e.g., lesbian, gay, bisexual, transgender, queer or questioning, intersex, and asexual). As different LGBTQ+ youth likely have unique experiences concerning their sexual and gender identities, as well as their online risk experiences, studying specific subgroups of the LGBTQ+ community might yield important insights not uncovered in our study. We suggest that future studies oversample from LGBTQ+ communities to achieve large enough sample sizes to detect medium to large size effects between subgroups within the LGBTQ+ community. While a strength of our research is examining the private online communications and risk experiences of LGBTQ+ and heterosexual youth, this means that our results should not be generalized to their public social media communications. Further, private communications via Instagram may also differ from those of other social media platforms. Therefore, we encourage future research to take into account different private versus public contexts, as well as diversify to study youths' experiences on other popular social media platforms. \edit{Ultimately, the landscape of adolescent mental health is likely influenced by a multitude of factors beyond private conversations. Future research should explore additional factors that play a role in shaping the mental well-being of LGBTQ+ youth, extending beyond the scope of their online risk experiences.}

\section{Conclusion}
The present study offers empirical evidence regarding the online risk experiences encountered by LGBTQ+ youth in private social media spaces. Furthermore, we examined how these interactions may impact the overall mental health of young adults. We conducted a mixed-method study to understand the severity and types of risks within online private messages experienced by LGBTQ+ youth. In doing so, we differentiate various types of risky and uncomfortable experiences of LGBTQ+ youth and show that LGBTQ+ youth face more severe risks in private spaces than their heterosexual peers. In addition, our findings highlight the relationship between sexual messages and Self-Harm behavior among both LGBTQ+ and heterosexual youth. In contrast, young people identifying as LGBTQ+ are more likely to Self-Harm due to harassment, whereas heterosexual youth are less affected. At the same time, an increased number of sexual messages corresponded to positive mental Well-Being among LGBTQ+ youth. Thematic analysis of DMs show that LGBTQ+ youth engage in risky conversations with strangers and frequently receive sexual messages and solicitations with inappropriate content. Our findings highlight the critical need for social media tools and support mechanisms to combat risks that take place within private online spaces and are particularly likely to negatively effect LGBTQ+ youth.

\begin{acks}
This research is supported in part by the U.S. National Science Foundation under grants \#IIP-2329976, \#IIS-2333207 and by the William T. Grant Foundation grant \#187941. Any opinions, findings, and conclusions or recommendations expressed in this material are those of the authors and do not necessarily reflect the views of the research sponsors. We would also like to thank all the participants who donated their data and contributed towards our research.
\end{acks}

\bibliographystyle{ACM-Reference-Format}
\bibliography{CHI2024}

\newpage
\onecolumn
\appendix

\label{Appendices}
\begin{appendices}

\section{\edit{Multivariate regression model}}
\label{appendix:a}
 

\begin{table*}[htbp]
  \footnotesize
  \caption{\edit{Unstandardized estimates for multivariate regression model (M4) examining the relationship between online risky messages (e.g., sexual messages, harassment, and self-injury) and individual Self-Harm and Injury Behaviors (ISAS), including Cutting (ISAS1), Severe Scratching (ISAS2), Biting (ISAS3)}}
  \label{ISAS1-3}
 \resizebox{\textwidth}{!} {
 \edit{
 \begin{tabular}{lccccccccc}
    \toprule
    & \multicolumn{3}{c}{\textbf{Cutting (ISAS1)}} & \multicolumn{3}{c}{\textbf{Severe Scratching (ISAS2)}} & \multicolumn{3}{c}{\textbf{Biting (ISAS3)}} \\
    \textit{Predictors} &  \textit{Estimates} &  \textit{Std. CI} & \textit{p}  &  \textit{Estimates} &  \textit{Std. CI} &  \textit{p} &  \textit{Estimates} &  \textit{Std. CI} &  \textit{p} \\
    \midrule
    (Intercept) & 1.116 & 0.791 - 1.441 & <0.00*** & 1.395 & 1.088 - 1.702 & <0.00*** & 1.267 & 0.965 - 1.568 & <0.00***\\
    \\
    Sexual Messages & 0.016 & -0.001 - 0.035 & 0.072 & 0.010 & -0.006 - 0.027 & 0.235 &  0.002 & -0.014 - 0.019 & 0.784\\
     \\
    Harassment & 0.004 & -0.010 - 0.018 &  0.558 & -0.004 & -0.018 - 0.009 & 0.529 & -0.000 & -0.013 - 0.013 & 0.952\\
    \\
    Self-Injury & 0.194 & -0.082 - 0.472 & 0.167 & 0.109 & -0.151 - 0.371 & 0.408 & -0.104 & -0.361 - 0.153 & 0.425\\
    \\
    \textbf{Sexual Identity (LGBTQ+)} & \textbf{0.836} & \textbf{0.345 -  1.327} & \textbf{0.000***} & \textbf{0.591} & \textbf{0.129 -  1.054} & \textbf{0.012*} & \textbf{0.622} & \textbf{1.666 - 1.077} & \textbf{ 0.007***}\\
    \\
    Sexual Messages x Sexual Identity (LGBTQ+) & -0.013  & -0.037 - 0.010 & 0.268 & -0.008  & -0.031 - 0.014 & 0.449 & -0.002  & -0.024 - 0.020 & 0.841\\
    \\
    Harassment x Sexual Identity (LGBTQ+)  & \textbf{0.031}  & \textbf{0.001 - 0.062} & \textbf{0.040\oneS} & 0.018  & -0.010 - 0.047 & 0.200 & 0.003  & -0.024 - 0.031 & 0.816\\
    \\
    Self-injury x Sexual Identity (LGBTQ+)  & -0.019  & -0.316 - 0.277 & 0.896 & -0.112  & -0.392 - 0.167 & 0.426 & 0.127  & -0.148 - 0.402 & 0.363 \\
    \\
  \midrule
  Observations & \multicolumn{3}{c}{173} & \multicolumn{3}{c}{173} & \multicolumn{3}{c}{173}\\
  $R^2$ / Adjusted $R^2$ & \multicolumn{3}{c}{0.263 / 0.232} & \multicolumn{3}{c}{0.101 / 0.063} & \multicolumn{3}{c}{0.105 / 0.068}\\
      \bottomrule 
  \\
  \multicolumn{9}{l}{Note.  *\textit{p} < .05; **\textit{p}<.01; ***\textit{p}<.001}
  \\
\end{tabular}}}
\end{table*}

\begin{table*}[htbp]
  \footnotesize
    \caption{\edit{Unstandardized estimates for M4 examining the relationship between online risky messages (e.g., sexual messages, harassment, and self-injury) and individual Self-Harm and Injury Behaviors (ISAS), including Banging or Hitting Self (ISAS4), Burning (ISAS5), Interfering with Wound Healing (ISAS6)}}
  \label{ISAS4-6}
 \resizebox{\textwidth}{!} {
 \edit{
 \begin{tabular}{lccccccccc}
    \toprule
    & \multicolumn{3}{c}{\textbf{Banging or Hitting Self (ISAS4)}} & \multicolumn{3}{c}{\textbf{Burning (ISAS5)}} & \multicolumn{3}{c}{\textbf{Interfering w/ Wound Healing (ISAS6)}} \\
    \textit{Predictors} &  \textit{Estimates} &  \textit{Std. CI} & \textit{p}  &  \textit{Estimates} &  \textit{Std. CI} &  \textit{p} &  \textit{Estimates} &  \textit{Std. CI} &  \textit{p} \\
    \midrule
    (Intercept) & 1.319 & 1.005 - 1.633 & <0.00*** & 1.055 & 0.856 - 1.254 & <0.00*** & 2.063 & 1.621 - 2.505 & <0.00***\\
    \\
    Sexual Messages & 0.015 & -0.002 - 0.033 & 0.086 & 0.003 & -0.007 - 0.014 & 0.508 &  0.013 & -0.011 - 0.038 & 0.282\\
     \\
    Harassment & -0.004 & -0.018 - 0.009 &  0.522 & 0.002 & -0.006 - 0.011 & 0.588 & -0.002 & -0.021 - 0.017 & 0.832\\
    \\
    Self-Injury & 0.090 & -0.177 - 0.358 & 0.506 & -0.029 & -0.199 - 0.139 & 0.729 & -0.123 & -0.500 - 0.253 & 0.517\\
    \\
    Sexual Identity (LGBTQ+) & \textbf{0.697} & \textbf{0.223 -  1.170} & \textbf{0.004**} & 0.201 & -0.098 -  0.501 & 0.187 & 0.368 & -0.298 - 1.035 &  0.277\\
    \\
    Sexual Messages x Sexual Identity (LGBTQ+) & -0.016  & -0.039 - 0.007 & 0.170 & 0.002  & -0.012 - 0.016 & 0.778 & -0.013  & -0.045 - 0.019 & 0.427\\
    \\
    Harassment x Sexual Identity (LGBTQ+)  & 0.021  & -0.007 - 0.050 & 0.152 & 0.006  & -0.011 - 0.025 & 0.468 & 0.0325  & -0.008 - 0.073 & 0.121\\
    \\
    Self-injury x Sexual Identity (LGBTQ+)  & -0.042  & -0.329 - 0.244 & 0.769 & 0.053  & -0.128 - 0.234 & 0.563 & 0.275  & -0.127 - 0.678 & 0.179 \\
    \\
  \midrule
  Observations & \multicolumn{3}{c}{173} & \multicolumn{3}{c}{173} & \multicolumn{3}{c}{173}\\
  $R^2$ / Adjusted $R^2$ & \multicolumn{3}{c}{0.129 / 0.092} & \multicolumn{3}{c}{0.078 / 0.039} & \multicolumn{3}{c}{0.090 / 0.025}\\
      \bottomrule 
  \\
  \multicolumn{9}{l}{Note.  *\textit{p} < .05; **\textit{p}<.01; ***\textit{p}<.001}
  \\
\end{tabular}}}
\end{table*}

\begin{table*}[htbp]
  \footnotesize
  \caption{\edit{Unstandardized estimates for M4 examining the relationship between online risky messages (e.g., sexual messages, harassment, and self-injury) and individual Self-Harm and Injury Behaviors (ISAS), including Carving (ISAS7), Rubbing Skin against Rough Surface (ISAS8), Pinching (ISAS9)}
  \label{ISAS7-9}}
 \resizebox{\textwidth}{!} {
 \edit{
 \begin{tabular}{lccccccccc}
    \toprule
    & \multicolumn{3}{c}{\textbf{Carving (ISAS7)}} & \multicolumn{3}{c}{\textbf{Rubbing Skin Against Rough Surface (ISAS8)}} & \multicolumn{3}{c}{\textbf{Pinching (ISAS9}} \\
    \textit{Predictors} &  \textit{Estimates} &  \textit{Std. CI} & \textit{p}  &  \textit{Estimates} &  \textit{Std. CI} &  \textit{p} &  \textit{Estimates} &  \textit{Std. CI} &  \textit{p} \\
    \midrule
    (Intercept) & 1.075 & 0.881 - 1.269 & <0.00*** & 1.364 & 1.056 - 1.671 & <0.00*** & 1.305 & 1.008 - 1.602 & <0.00***\\
    \\
    Sexual Messages & \textbf{0.113} & \textbf{0.000 - 0.022} & \textbf{0.048\oneS} & 0.008 & -0.008 - 0.025 & 0.345 & 0.008 & -0.007 - 0.025 & 0.304\\
     \\
    Harassment & -0.003 & -0.012 - 0.004 &  0.393 & -0.007 & -0.213 - 0.006 & 0.273 & -0.002 & -0.016 - 0.010 & 0.658\\
    \\
    Self-Injury & -0.014 & -0.180 - 0.150 & 0.860 & -0.141 & -0.403 - 0.120 & 0.289 & 0.009 & -0.243 - 0.262 & 0.940\\
    \\
    \textbf{Sexual Identity (LGBTQ+)} & -0.076 & -0.368 -  0.216 & 0.608 & 0.055 & 0.408 -  0.518 & 0.815 & 0.244 & -0.203 - 0.691 & 0.283\\
    \\
    Sexual Messages x Sexual Identity (LGBTQ+) & -0.006  & -0.020 - 0.008 & 0.410 & -0.004  & -0.027 - 0.018 & 0.686 & 0.000  & -0.021 - 0.022 & 0.932\\
    \\
    Harassment x Sexual Identity (LGBTQ+)  & \textbf{0.023}  & \textbf{0.005 - 0.041} & \textbf{0.012\oneS} & \textbf{0.034}  & \textbf{0.005 - 0.062} & \textbf{0.019*} & 0.033  & 0.006 - 0.061 & 0.017\\
    \\
    Self-injury x Sexual Identity (LGBTQ+)  & 0.080  & -0.096 - 0.257 & 0.372 & 0.131  & -0.149 - 0.411 & 0.356 & -0.311  & -0.311 - 0.230 & 0.767 \\
    \\
  \midrule
  Observations & \multicolumn{3}{c}{173} & \multicolumn{3}{c}{173} & \multicolumn{3}{c}{173}\\
  $R^2$ / Adjusted $R^2$ & \multicolumn{3}{c}{0.093 / 0.055} & \multicolumn{3}{c}{0.070 / 0.030} & \multicolumn{3}{c}{0.122 / 0.084}\\
      \bottomrule 
  \\
  \multicolumn{9}{l}{Note.  *\textit{p} < .05; **\textit{p}<.01; ***\textit{p}<.001}
  \\
\end{tabular}}}
\end{table*}

\begin{table*}[htbp]
  \footnotesize
  \caption{\edit{Unstandardized estimates for M4 examining the relationship between online risky messages (e.g., sexual messages, harassment, and self-injury) and individual Self-Harm and Injury Behaviors (ISAS), including Sticking Self with Needles (ISAS10), Pulling Hair (ISAS11), Swallowing Dangerous Substances (ISAS12)}}
  \label{ISAS10-12}
 \resizebox{\textwidth}{!} {
 \edit{
 \begin{tabular}{lccccccccc}
    \toprule
    & \multicolumn{3}{c}{\textbf{Sticking Self w/ Needles (ISAS10)}} & \multicolumn{3}{c}{\textbf{Pulling Hair (ISAS11)}} & \multicolumn{3}{c}{\textbf{Swallowing Dangerous Substances (ISAS12)}} \\
    \textit{Predictors} &  \textit{Estimates} &  \textit{Std. CI} & \textit{p}  &  \textit{Estimates} &  \textit{Std. CI} &  \textit{p} &  \textit{Estimates} &  \textit{Std. CI} &  \textit{p} \\
    \midrule
    (Intercept) & 1.055 & 0.827 - 1.283 & <0.00*** & 1.366 & 1.038 - 1.693 & <0.00*** & 1.023 & 0.874 - 1.173 & <0.00***\\
    \\
    \textbf{Sexual Messages} & 0.011 & -0.000 - 0.024 & 0.067 & \textbf{0.020} & \textbf{0.002 - 0.039} & \textbf{0.026\oneS} &  0.001 & -0.007 - 0.009 & 0.770\\
     \\
    Harassment & 0.000 & -0.009 - 0.010 &  0.970 & 0.003 & -0.011 - 0.017 & 0.664 & 0.007 & 0.000 - 0.013 & 0.034*\\
    \\
    Self-Injury & -0.039 & -0.234 - 0.154 & 0.685 & -0.066 & -0.345 - 0.212 & 0.636 & -0.019 & -0.147 - 0.107 & 0.758\\
    \\
    \textbf{Sexual Identity (LGBTQ+)} & -0.055 & 0.399 -  0.288 & 0.751 & 0.281 & -0.211 -  0.775 & 0.261 & 0.180 & -0.0448 - 0.406 & 0.116\\
    \\
    Sexual Messages x Sexual Identity (LGBTQ+) & 0.005  & -0.011 - 0.022 & 0.505 & -0.017  & -0.041 - 0.007 & 0.164 & 0.000  & -0.010 - 0.011 & 0.960\\
    \\
    Harassment x Sexual Identity (LGBTQ+)  & 0.019  & 0.001 - 0.041 & 0.067 & 0.020  & -0.009 - 0.051 & 0.182 & 0.001  & -0.012 - 0.015 & 0.815\\
    \\
    Self-injury x Sexual Identity (LGBTQ+)  & 0.084  & -0.123 - 0.293 & 0.422 & 0.119  & -0.179 - 0.417 & 0.432 & 0.048  & -0.088 - 0.184 & 0.488 \\
    \\
  \midrule
  Observations & \multicolumn{3}{c}{173} & \multicolumn{3}{c}{173} & \multicolumn{3}{c}{173}\\
  $R^2$ / Adjusted $R^2$ & \multicolumn{3}{c}{0.137 / 0.100} & \multicolumn{3}{c}{0.084 / 0.039} & \multicolumn{3}{c}{0.094 / 0.056}\\
      \bottomrule 
  \\
  \multicolumn{9}{l}{Note.  *\textit{p} < .05; **\textit{p}<.01; ***\textit{p}<.001}
  \\
\end{tabular}}}
\end{table*}

\end{appendices}
\end{document}